\documentclass[fleqn,usenatbib,useAMS]{mnras}

\usepackage{graphicx}	
\usepackage{amsmath,nicefrac}	
\usepackage{amssymb}	
\usepackage{multicol}        
\usepackage{bm}		
\usepackage{pdflscape}	

\newcommand{\cms}{\,cm\,s$^{-1}$} 
\newcommand{\ms}{\,m\,s$^{-1}$} 

\usepackage[T1]{fontenc}
\usepackage{ae,aecompl}

\usepackage{newtxtext,newtxmath}

\usepackage{orcidlink}

\title[A. Anna John et. al]{{Granulation on a quiet K dwarf: HD\,166620\\
I. Spectral signatures as a function of line-formation temperature \hfill}}
\author[]{
A. Anna John$^{\orcidlink{0000-0002-1715-6939}\,1,2,3}$\thanks{Contact e-mail: \href{mailto:a.a.john@bham.ac.uk}{a.a.john@bham.ac.uk}},
K. Al Moulla$^{\orcidlink{0000-0002-3212-5778}\,4,5}$\thanks{SNSF Postdoctoral Fellow}, N. K. O'Sullivan$^{\orcidlink{0000-0002-0509-3627}\,6}$,
J. Fitzpatrick$^{\orcidlink{0009-0001-5149-4923}\,7}$,
A. Collier Cameron$^{\orcidlink{0000-0002-8863-7828}\,2,3}$,
\newauthor B. S. Lakeland$^{1}$,  M. Cretignier$^{6}$, A. Mortier$^{\orcidlink{0000-0001-7254-4363}\,1}$\thanks{UKRI Future Leaders Fellow}, Tim Naylor$^{\orcidlink{0000-0002-0506-8501}\,7}$, Joe Llama$^{8}$, S. Aigrain$^{6}$, C. Hartogh $^{2,3}$, \newauthor S. Dalal $^{7}$,  H. M. Cegla$^{9,10}$\textcolor{blue}{\footnotemark[3]}
, C. A. Watson\orcidlink{0000-0002-9718-3266}$^{11}$, X. Dumusque$^{5}$, A. F. Martínez Fiorenzano\orcidlink{0000-0002-4272-4272}$^{12}$
\\
$^{1}$School of Physics \& Astronomy, University of Birmingham, Edgbaston, Birmingham B15 2TT, UK\\
$^{2}$SUPA, School of Physics \& Astronomy, University of St Andrews, North Haugh, St Andrews, KY169SS, UK\\
$^{3}$Centre for Exoplanet Science, University of St Andrews, North Haugh, St Andrews, KY169SS, UK\\
$^{4}$Instituto de Astrofísica e Ciências do Espaço, Universidade do Porto, CAUP, Rua das Estrelas, 4150-762 Porto, Portugal\\
$^{5}$Observatoire Astronomique de l’Université de Genève, Chemin Pegasi 51, 1290 Versoix, Switzerland\\
$^{6}$Denys Wilkinson Building, Department of Physics, University of Oxford, OX1 3RH, UK\\
$^{7}$Department of Physics and Astronomy, University of Exeter, Exeter EX4 4QL, UK\\
$^{8}$Lowell Observatory, 1400 W Mars Hill Road, Flagstaff, AZ, 86001, USA\\
$^{9}$Department of Physics, University of Warwick, Gibbet Hill Road, Coventry CV4 7AL, UK\\
$^{10}$Centre for Exoplanets and Habitability, University of Warwick, Gibbet Hill Road, Coventry CV4 7AL, UK\\
$^{11}$Astrophysics Research Centre, School of Mathematics and Physics, Queen’s University Belfast, Belfast, BT7 1NN, UK\\
$^{12}$Fundación Galileo Galilei - INAF, Rambla José Ana Fernandez Pérez 7, E-38712 Breña Baja, Tenerife, Spain
}

\date{Accepted 02 September 2025; for publication in MNRAS}

\pubyear{2025}

\begin{document}
\label{firstpage}
\pagerange{\pageref{firstpage}--\pageref{lastpage}}
\maketitle

\begin{abstract}
As Radial velocity (RV) spectrographs reach unprecedented precision and stability below 1 \ms, the challenge of granulation in the context of exoplanet detection has intensified. Despite promising advancements in post-processing tools, granulation remains a significant concern for the EPRV community.
We present a pilot study to detect and characterise granulation using the  High-Accuracy Radial-velocity Planet Searcher for the Northern hemisphere  (HARPS-N) spectrograph. We observed HD\,166620, a K2 star in the Maunder Minimum phase, intensely for two successive nights, expecting granulation to be the dominant nightly noise source in the absence of strong magnetic activity. Following the correction for a newly identified instrumental signature arising from illumination variations across the CCD, we detected the granulation signal using structure functions and a one-component Gaussian Process (GP) model. The granulation signal exhibits a characteristic timescale of $43.65^{+16.9}_{-14.7}$ minutes, within one $\sigma$, and a standard deviation of $22.9^{+0.83}_{-0.72}$ \cms, within three $\sigma$ of the predicted value.
By examining spectra and RVs as a function of line formation temperature, we investigated the sensitivity of granulation-induced RV variations across different photospheric layers. We extracted RVs from various photospheric depths using both the line-by-line (LBL) and cross-correlation function (CCF) methods to mitigate any extraction method biases. Our findings indicate that granulation variability is detectable in both temperature bins, with the cooler bins, corresponding to the shallower layers of the photosphere, aligning more closely with predicted values. 

\end{abstract}




\begingroup
\let\clearpage\relax
\endgroup
\newpage

\section{Introduction}

High-resolution spectroscopy and Doppler Radial velocities (RVs) are among the main workhorses in exoplanet research. Yet, the challenge lies in achieving RV precision of 10 \cms required for identifying Earth-mass planets in the Habitable Zone of Sun-like stars (Earth-twins), due to intrinsic stellar variability. Despite the technological feasibility of current extremely precise radial velocity (EPRV) instruments to attain $<$30\,\cms precision, stellar variability remains a significant hindrance. 


Stellar variability encompasses a range of astrophysical processes \citep[see fig. A-3 from][]{Crass2021} that impact the RV of stars. This ranges from short timescale processes like oscillations, and granulation, which occur on minutes to hour timescales \citep[][]{Gray2009, Meunier2015}, to longer-lived structures such as supergranulation, which evolves over several days, and magnetic activity features like spots, faculae, and plages that persist for days to months.
Current stellar activity mitigation methods, encompassing time and wavelength domain approaches, have enabled the extraction of planet signals from RV data with amplitudes of 1-2\,\ms, amid stellar variability \citep[eg:][]{Haywood2014, Rajpaul2015, Faria2018, ACC2021, deBeurs2022, Klein2022}. However, for most Earth-sized exoplanets detected by TESS, \textit{Kepler}, and soon PLATO, more advanced methods are required to convert mass upper limits into precise absolute measurements. Novel post-processing tools \citep[eg:][]{Cretignier2021, AnnaJohn2023, Klein2024} have shown promise to further mitigate non-planetary signals below 1 \ms on stellar targets at the spectrum level. Nonetheless, all these attempts appear to be capped at $\sim$50\,\cms, a level comparable to RV variations induced by granulation and supergranulation \textcolor{black} {on stellar targets. Recently, \citet{ZhaoY2024A} demonstrated advanced activity mitigation, achieving a notable threshold of 20 \cms in solar data. }

Convection in low-mass stars drives oscillations caused by acoustic $p$-modes and the formation of granulation patterns on the stellar surface. These patterns are characterised by up-flowing hot plasma and down-flowing cool plasma, resulting in blue-shifted granules and red-shifted inter-granular lanes. Granules, compared to intergranular lanes, generally cover a larger surface area and are more luminous due to the hot plasma, leading to a significant contribution to the stellar flux and a net convective blueshift \citep[eg:][]{Dravins1981, Dalal2023}. This process causes time-variable asymmetries in stellar absorption lines and is well-studied in the Sun, where granules are visible. As granules on the Sun evolve over their lifetimes of 5-6 minutes, the changing ratio of granules to intergranular lanes leads to continual variations in stellar line asymmetries and net RV shifts \citep[eg:][and references therein]{Cegla2019}. Using 3D magneto-hydrodynamic (MHD) simulations and radiative transfer modelling, \citet{Sulis2020, Cegla2013, Cegla2018, Cegla2019b} quantified the RV signature of granulation, revealing its correlation with line asymmetry. This signature adds significant complexity to the detection and characterisation of low-amplitude RV shifts induced by Earth twins \citep{Dravins1986}. Granulation primarily affects stellar spectra in three ways: (1) broadening individual lines \citep{Gray2009} (2) disturbing the line symmetry \citep[eg:][]{Lohner2018, Gray2009}, and (3) inducing absolute convective blueshift \citep[eg:][]{Gray2009, Reiners2016}. These effects tend to be more pronounced in weaker spectral lines. \citep[eg:][]{Meunier2017, Reiners2016}. Across the stellar disc, the upflows and downflows largely cancel out, resulting in net variability of 0.1–1\,\ms for Sun-like stars \citep{Schrijver2000}. \citet{Lakeland2024} identified granulation as one of the primary sources of RV variability during Solar minimum. They measured the granulation-induced RV variability in the Sun, reporting an amplitude of $\sim$ 40\,\cms \citep[eg:][]{ACC2019, Sulis2020, Lakeland2024}.

Stellar granulation, with its average timescale of approximately 8 hours, provides valuable insights into stellar physics but significantly hampers the detection of low-mass exoplanets. To date, exoplanet-focused RV observations have primarily attempted to alleviate the impact of granulation by adopting exposure times that average out their short timescale variability \citep[][]{Dumusque2011}. This involves averaging two or three exposures of at least 900\,s taken $>$2 hours apart within a night. This approach has been a long-standing standard at most EPRV facilities. However, in a study using simulations of granulation in both photometric and RV time series covering a full solar cycle, \citet{Meunier2015} demonstrated that this averaging strategy is more effective at reducing root-mean-squared (RMS) granulation noise compared to consecutive observations—though not to the level reported by \citet{Dumusque2011}. \citet{Meunier2015} found that granulation-induced RV signals remain significant, even after averaging, with a root-mean-squared (RMS) value of approximately 0.4 \ms. Therefore, it is crucial to understand and quantify granulation in stellar targets to achieve the desired RV precision, rather than relying solely on observational strategies that attempt to `average' out such signals.

\begin{figure}
    \centering
    \includegraphics[width=1\linewidth]{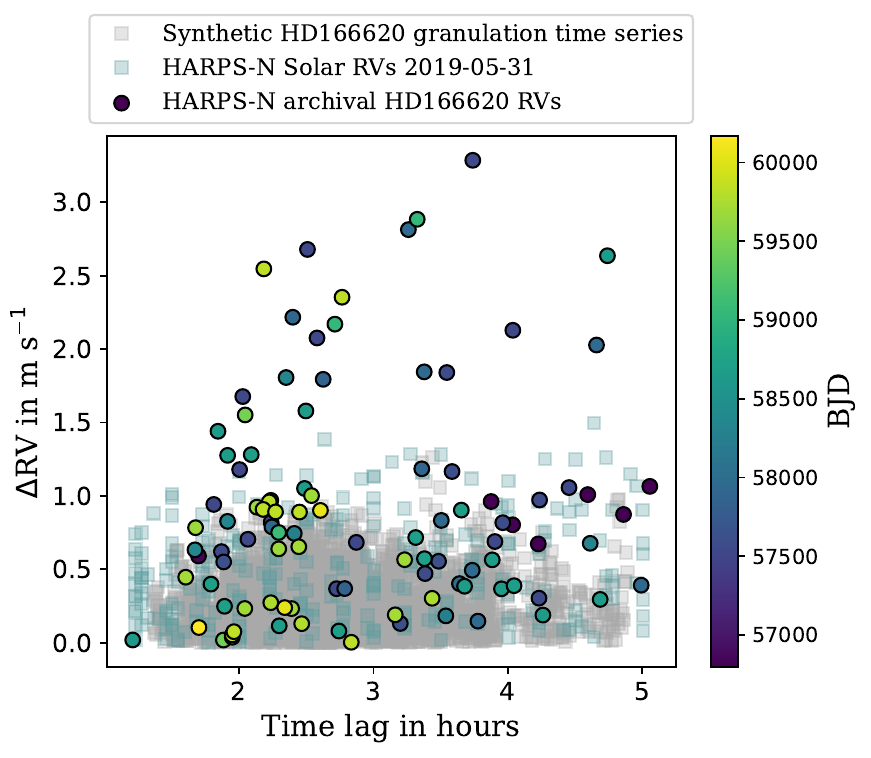}
    \caption{ 141 intra-night pairs of 900-s HARPS-N RPS observations of HD\,166620 secured between 100 minutes and 5 hours apart since 2012. These show statistically significant RMS velocity scatter, expected to be due to granulation, that increases with the lag between observations. The bifurcation in the RMS distribution is concerning, as we cannot establish any time dependency if it is instrumental, demanding immediate investigation. \textcolor{black}{The grey squares in the background shows comparable pairs from the synthetic granulation time series mentioned in Section \ref{sec:SF_GP_simulations}. The teal squares represent corresponding pairs from solar HARPS-N data obtained on a spot-free clear day during activity minimum.}}
    \label{fig:RPS intranight}
\end{figure}

\citet{Sulis2023} recently observed two bright stars (F \& G spectral types) using ESPRESSO and CHEOPS simultaneously, aiming to link the photospheric and spectroscopic signatures of granulation. Although they could detect and characterise granulation precisely in the F star, they could not do so for the G star. The challenge stemmed from the difficulty in establishing a correlation between the stellar signal and the CCF line shape variations, attributed to SNR-dependent fluctuations. They showed that the temporal binning of the RV time series is limited in mitigating the granulation signal, which remains correlated over long timescales, agreeing with the conclusions from \citet{Meunier2015}. 

This paper presents our pioneering attempt to study granulation by probing stellar surface convection in a K-type star through direct observational investigations from the highly wavelength-stabilised spectrograph, HARPS-N. Our pilot study focuses on detecting and characterising granulation signatures from spectroscopic observations of the magnetically quiet, bright K-dwarf, HD\,166620. We aim to quantify the amplitudes and timescales of granulation, investigate the sensitivity of different photospheric depths to granulation signatures, and provide EPRV-level diagnostics of how granulation manifests in stellar spectra and RV data.
The results will be crucial for refining techniques to mitigate stellar variability, identifying instrument capabilities, and constraining and informing existing models.

The manuscript is organised as follows: Section \ref{sec:obs} outlines the motivation for this research, target selection, and observational strategy. We will discuss the instrumental systematics discovered during this study and the correction strategies adopted in Section \ref{sec:instrumental}. In Section \ref{sec:analysis}, we outline the process of estimating the spectral line formation temperature using spectral synthesis. Also, we detail the extraction of formation temperature-dependent RVs to evaluate the sensitivity of granulation-induced variations at different photospheric depths. We present the detection and characterisation of granulation signals from various temperature-dependent RV time-series using Gaussian process (GP) models in Section \ref{sec:results}, followed by the discussion and conclusions.

\begin{figure*}
    \centering
    \includegraphics[width=0.92\linewidth, trim=1cm 0 0 0]{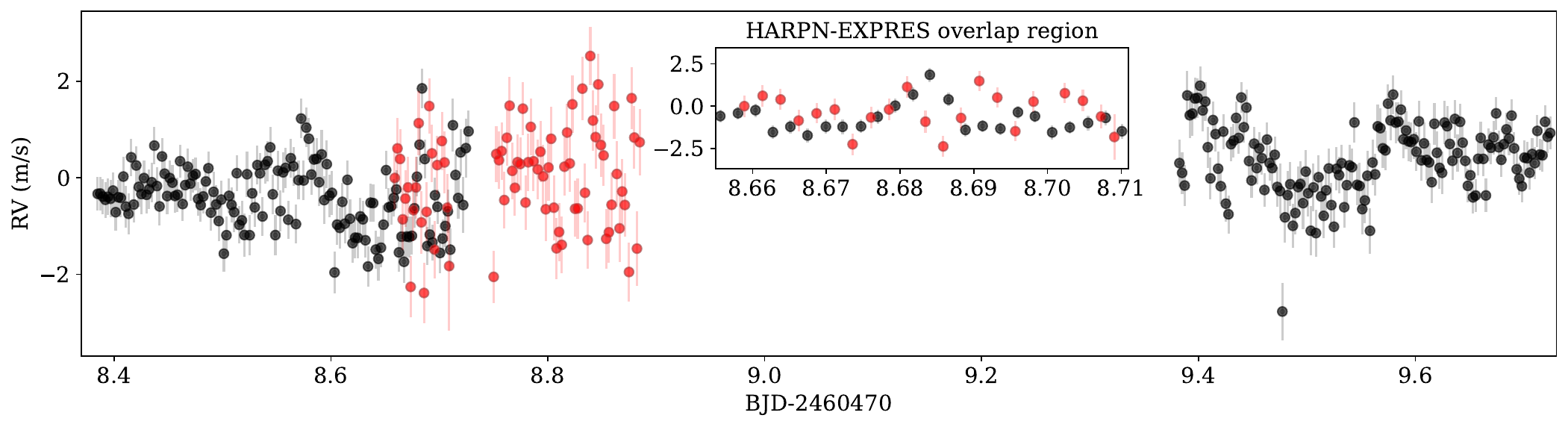}
    \includegraphics[width=1\textwidth]{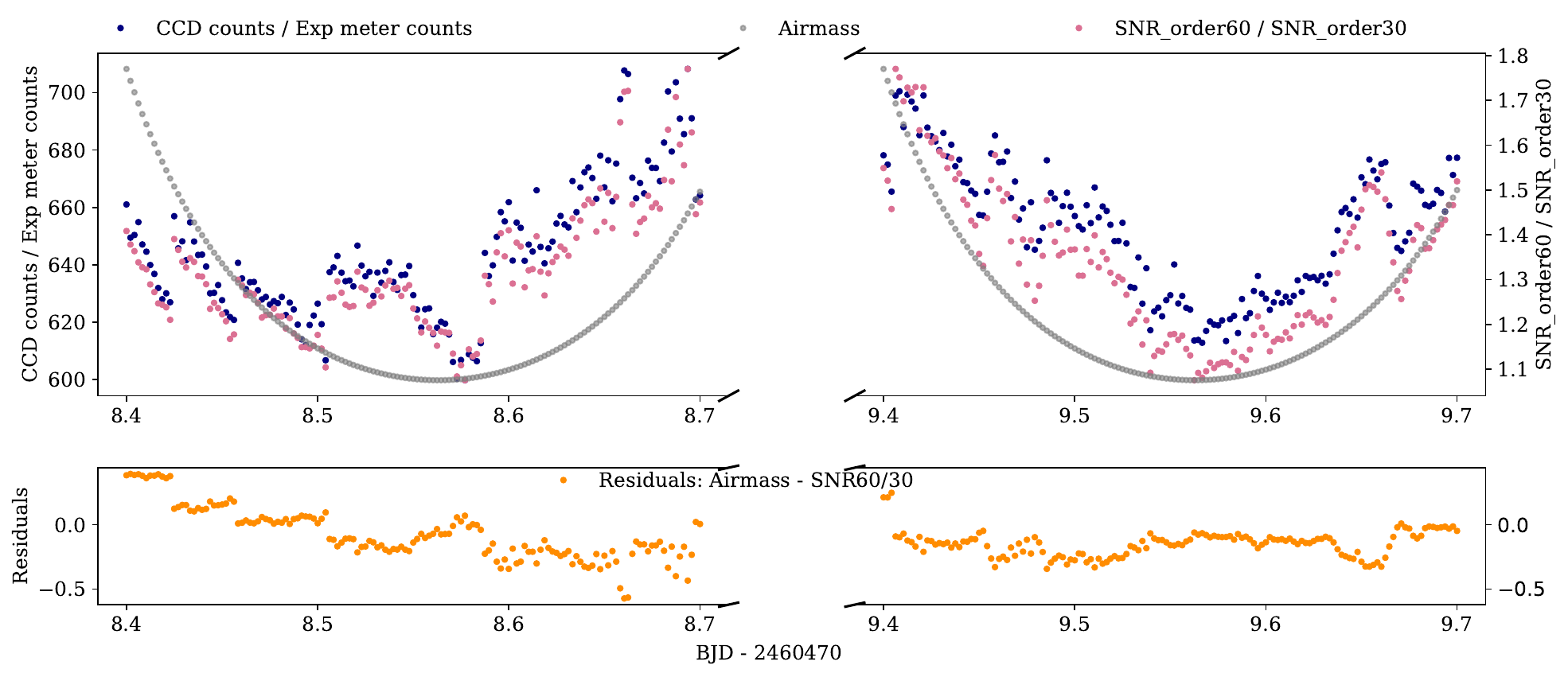} 
    \caption{\textbf{Top:} The contemporaneously acquired HARPS-N (black) and EXPRES (red) radial-velocity observations of HD\,166620.
    \textbf{Middle:} Diagnostic ratios CCD counts/exposure-meter counts (dark blue) and SN60/SN30 (dark pink), plotted against Julian date of the HARPS-N observations. Both metrics display a striking resemblance, with measured correlation coefficients of 0.99 and 0.98 for nights 1 and 2, respectively. The airmass is shown in grey. 
    \textbf{Bottom:} The residuals after subtracting the airmass from the ratios, revealing distinct structures. This suggests that the observed patterns are not solely attributable to the colour effect.}
    \label{fig:DiagnosticRatios}
\end{figure*}

\section{Target and Observations}
\label{sec:obs}
\subsection{HD\,166620: The quiet, Maunder Minimum analogue}

 The Maunder Minimum refers to a period from 1645 to 1715 when the Sun experienced a prolonged minimum in sunspot activity, with fewer than fifty sunspots observed during this period, compared to the usual 40,000 to 50,000 \citep{JohnEddy1976}. HD\,166620, a K2V type RV standard star with a magnitude of 6.38 \citep{Motalebi2015}, has been extensively monitored for its kinematic, spectroscopic and photometric properties. Recent studies \citep[][]{Luhn2022, Baum2022, AnnaJohn2023} have definitively confirmed that this star entered a prolonged activity minimum in the early 2000s and has remained in this state since then, making HD\,166620 the only known Maunder Minimum analogue.  This conclusion is based on five decades of chromospheric activity monitoring by the Mount Wilson program \citep{Duncan1991} and continued observations at Keck as part of the California Planet Search programme \citep[][]{Wright2004, Isaacson2010}, along with S-index measurements from HARPS-N \citep[see fig. 3 of][]{AnnaJohn2023}. \cite{AnnaJohn2023} updated the stellar parameters of HD\,166620 (see table 2 of that work) using 12 years of HARPS-N Rocky Planet Search (RPS) observations and found no significant planetary signals in the RV data, down to $\sim$60 \cms. HD\,166620, being in a grand minimum activity state for the last two decades, presents as an excellent candidate for this study as we expect granulation to be the dominant source of astrophysical noise in the absence of significant magnetic activity.

Given HD\,166620 is cooler and less massive \citep[4989\,K, 0.77\,M$_\odot$;][]{AnnaJohn2023} than the Sun, with a $\log{g}$ of $4.55^{+0.12}_{-0.01}$, models \citep{Dalal2023} predict a granulation RMS lower than that of the Sun ($<$20\,\cms, as opposed to the $<$40\,\cms for the Sun), potentially falling below the detection threshold of HARPS-N. The RPS program's observing strategy to mitigate granulation effects, based on \cite{Dumusque2011}, renders the 1167 archival observations unsuitable for fully comprehending granulation phenomena. However, the 141 intra-night pairs of 900-s RPS observations secured between 100 minutes and 5 hours apart (obtained to average out the intra-night granulation noise) show a detectable RMS scatter around 50\,\cms (Fig.~\ref{fig:RPS intranight}). Fig.~\ref{fig:RPS intranight} reveals a wedge-shaped pattern in the RMS distribution, characterised by a clear bifurcation. This suggests the presence of one or more sources of correlated noise, to which granulation is likely to be a significant, though perhaps not dominant, contributor. Additionally, there remains the possibility of unidentified instrumental systematics. \textcolor{black}{Simulated granulation time series for HD\,166620 (discussed in Sec. \ref{sec:SF_GP_simulations}) and HARPS-N solar observations taken on a clear, spot-free day further hint at a non-astrophysical origin for the upper branch, as both datasets exclusively trace the lower branch in Fig.~\ref{fig:RPS intranight}.} Investigating the source of this dispersion is crucial for achieving the 10\,\cms precision required for detecting Earth-twins. This motivated the strategically dense observations described in the next section.

\subsection{Contemporaneous HARPS-N and EXPRES observations}

HD\,166620 was observed continuously throughout the nights of 2024 June 16 and 17, using HARPS-N \citep{Cosentino2012}, and the EXtreme PREcision Spectrometer \citep[EXPRES;][]{Jurgenson2016}(EXPRES). HARPS-N is mounted at the f/11 Nasmyth-B focal station of the 3.5-m Telescopio Nazionale Galileo (TNG) on La Palma, and the EXPRES  is mounted on the 4.3-m Lowell Discovery Telescope (LDT) at Lowell Observatory in Northern Arizona. 
These spectrographs feature resolutions of R\,=\,115,000 \citep{Cosentino2012} and 137,500 \citep{Jurgenson2016}, respectively. Each night's HARPS-N science observations comprised 145 exposures, each lasting 180-s and achieving a median signal-to-noise ratio (SNR) of $\sim$180. The overall cadence, including the CCD readout time of 25s, was 205s per exposure. Over two nights, we obtained 290 (17 per hour, 145 per night) observations for HD\,166620, at sub-arcsecond seeing. The RVs were extracted using version 3.0.1 of the ESPRESSO Data Reduction System (DRS) pipeline, optimised for the HARPS-N instrument \citep{Dumusque2021}.  EXPRES observed HD\,166620 contemporaneously (64 x 170~s exposures, with fixed integration time) on the first night of the granulation campaign (16 June 2024). The target was unobservable on the second night due to smoke at the telescope site in Arizona. The EXPRES observations were reduced using the standard EXPRES pipeline (Petersburg 2020), and used ThAr and LFC  calibration sources and were also run through the Excalibur pipeline for optimal wavelength calibration \citep{Petersburg2020, Zhao2021}. 

HARPS-N observations overlapped with EXPRES observations for 1-1.5 hours, allowing for direct comparison (top panel of Fig.~\ref{fig:DiagnosticRatios}). The high-frequency and low-frequency RV variations observed by both telescopes show an adequate match in the region of overlapped observations. 

\subsection{Instrumental systematics}
\label{sec:instrumental}


HARPS-N is fed by a pair of optical fibres, whose octagonal cross-section has been shown to have scrambling properties superior to other designs \citep{Chazelas2010}. The fibre diameter projects to 1 arc second in the telescope focal plane. Since HD\,166620 is a bright target requiring high RV precision, the light from the target passed through fibre A, while the auxiliary fibre B was illuminated with light from the Fabry-Perot source to actively monitor instrument drift. The spectra of both fibres were recorded simultaneously on the CCD. The DRS computes the cross-correlation function \citep[CCF;][]{Baranne1996,Pepe2002} between the observed stellar spectrum and a precomputed weighted line mask indicating the central wavelengths and relative RV contributions of many expected spectral lines in the relevant wavelength range; specifically, the DRS uses the masks provided by the ESPRESSO pipeline, of which the K2 mask is applied for the CCFs of HD\,166620. The RV at each epoch is thereafter obtained by measuring the centre of the Gaussian fitted to the CCF.

The seeing during the run was exceptionally sharp. The Differential Image-Motion Monitor (DIMM) located near the TNG dome recorded seeing measurements around 0.6 arcsec for much of both nights. Since this is significantly smaller than the fibre entrance aperture, we monitored the integrated images from the camera and the tip-tilt auto guider. Minor changes were noted in the morphology of the halo recorded of the PSF around the fibre entrance aperture throughout both nights.

The HARPS-N RVs extracted by the DRS are plotted against Julian date in the black points shown in the upper panel of Fig.~\ref{fig:DiagnosticRatios}. They show unexpectedly strong RV variability on timescales of a few hours, with a peak-to-peak amplitude of $\sim$3.5\,\ms. 
This was considerably greater than the amplitude we anticipated for the granulation signal of a quiet K dwarf star. 

 We investigated the possibility of an environmental or instrumental contribution to the RV signal using metadata from the file headers. In particular, we examined the SNR in echelle order 60 (SN60) to that in order 30 (SN30), and the ratio of the total counts recorded in the extracted 2-dimensional echelle spectra to the integrated counts recorded by the exposure meter. As Fig.~\ref{fig:DiagnosticRatios} shows, the two ratios bore a powerful resemblance. The overarching `U'-shaped trend arises from the well-documented colour effect of airmass, where differential extinction affects light from varying wavelengths. Nonetheless, this effect does not entirely account for the observed structure. This is particularly evident in the residuals, which dominate the radial velocities extracted from the cooler regions of the spectra (Fig \ref{Day1steps}), as detailed in Section \ref{sec:temp_formalism}.

We thus investigated the possibility that seeing-related instability in the fibre injection at the spectrograph entrance led to a time-dependent, chromatic light loss. 

HARPS-N has a triple-pass white-pupil design, meaning that light emerging from the fibre entrance is reflected from the collimator to the echelle grating and exposure meter, then via the collimator to the cross-disperser, and finally returning via the collimator to the camera. The exposure meter is located between the two halves of the echelle grating, considerably upstream of the CCD detector in the light path through the instrument. If we suppose that changes in seeing and guiding fluctuations cause small variations in the direction and/or focal ratio of the beam emerging from the fibre, then the illumination of the collimator would change. As a consequence, this would affect the total flux reaching the CCD as well as its spatial distribution. Such variations could propagate into the RV measurements through changes in the relative weighting of the photon counts per pixel in different areas of the detector.
\begin{figure}
    \centering
    \includegraphics[width=0.95\linewidth]{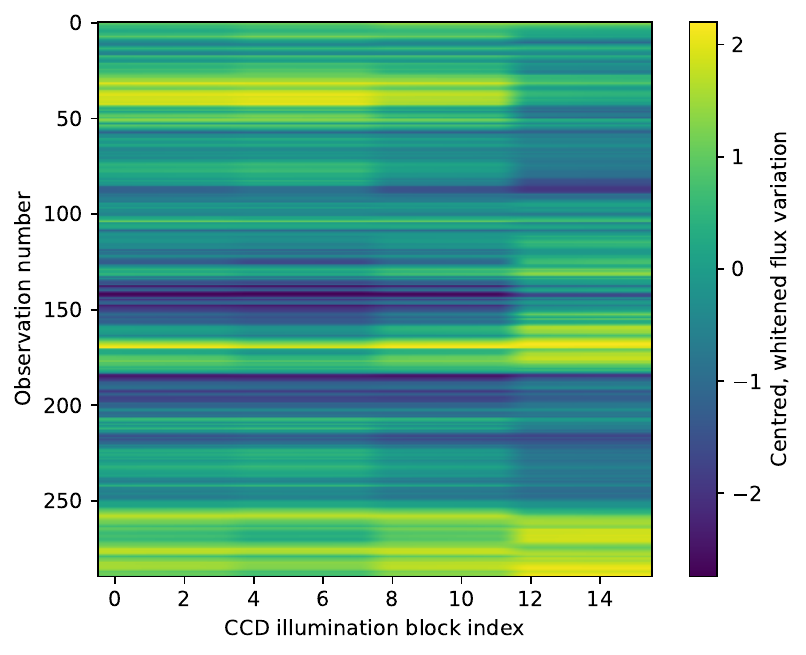}
    \caption{Density plot showing relative flux variations in the 4x4 array of sub-regions used to characterise the CCD illumination. The 16 columns show the centred, whitened flux variations in the blocks.}
    \label{fig:IllumArray}
\end{figure}
\begin{figure}
    \centering
    \includegraphics[width=0.95\linewidth]{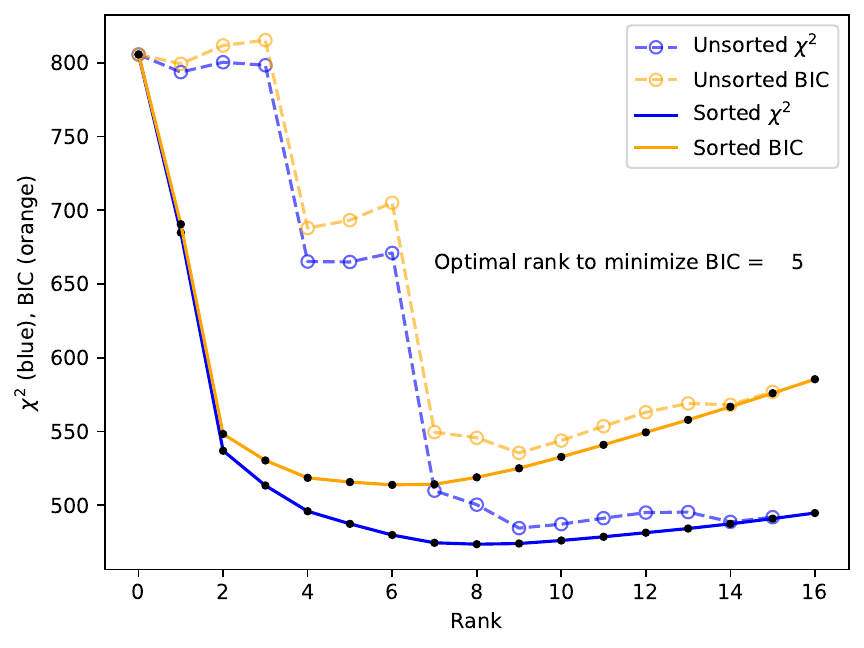}
    \caption{Optimal re-ordering of the columns of the left-singular matrix $\bm{U}$ shows the $\chi^2$ and BIC of the fit to the radial velocity as a function of the rank $k$ of the SVD reconstruction from the CCD illumination variations. This illustrates that the 4th and 7th principal components of the CCD illumination variations project most strongly onto the radial velocity.}
    \label{fig:rearrange_U}
\end{figure}
\begin{figure}
    \centering
    \includegraphics[width=0.7\linewidth,angle=270]{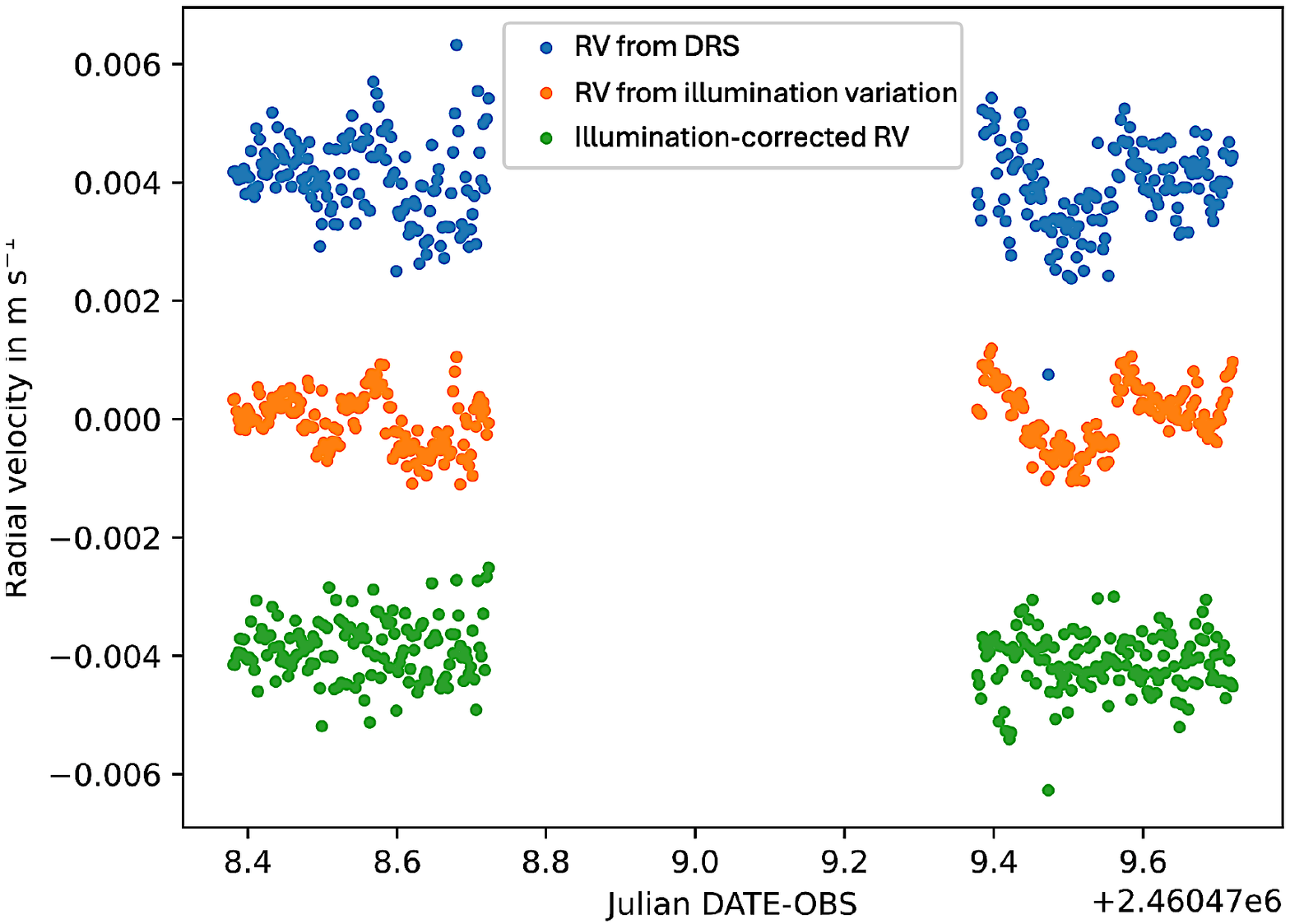}
    \caption{HARPS-N RVs are plotted as a function of Julian date of observation. The upper (blue) markers trace the velocities produced by the DRS. The middle (orange) sequence shows the projection of the RVs into a rank-4 SVD representation of the CCD illumination variations arising from fibre-injection instability. The bottom (green) markers trace the residual variation containing the astrophysical signal plus photon noise. The RVs are plotted with an offset for better illustration.}
    \label{fig:CCD_illum_correction}
\end{figure}
\begin{figure*}
    \centering
    \includegraphics[width=0.85\linewidth]{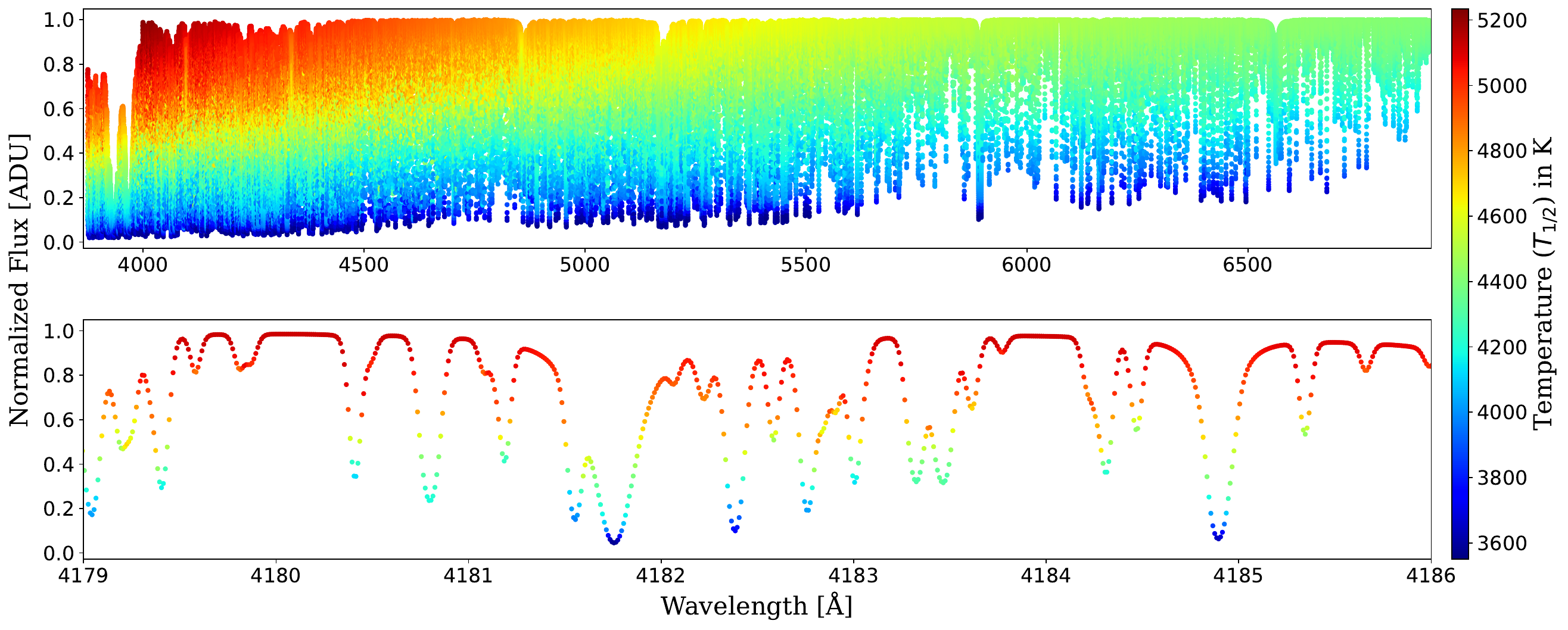}
    \caption{\textit{Top}: Normalised synthetic flux spectrum for K2, the spectral type of HD\,166620, colour-coded by the average formation temperature and interpolated onto the HARPS-N wavelengths after being corrected for the BERV and systematic velocity of HD\,166620. \textit{Bottom}: A zoomed-in wavelength window.}
    \label{fig:spectralwindow}
\end{figure*}
\label{sec:Tform}

Some previous investigations regarding fibre injection systematics were performed on the SOPHIE spectrograph \citep{Bouchy2013}. \citet{Bouchy2013} showed that one characteristic signature of such systematics is a velocity shift along the main order direction, with a gradient shift in amplitude from left-to-right\footnote{Note that, \citet{Bouchy2013} characterized the issue in the circular fibres, while in our case the issue was also detected on the octagonal fibres.}. To see if a similar result was obtained here, we extracted the line-by-line \citep[LBL;][]{Dumusque2018} RVs of the S2D spectra corrected by YARARA \citep{Cretignier2021}. The parallel investigation (detailed in Appendix~\ref{sec:LBL}) confirmed that the LBL RVs extracted from different physical regions of the detector showed distinctive patterns of variability whose form was similar to that seen in the velocities measured by the DRS, but with different amplitudes. It illustrates one more time how LBL RVs may be used to disentangle planetary signals from stellar or instrumental effects, as pointed out in \citet{Cretignier2023}.  

\textcolor{black}{While \citet{Meunier2024} also observed a $\sim$ 2\ms signal with a $\sim$ 200-day period in solar RV and bisector measurements, attributed to periodic detector warm-ups, the signal we report here arises from fibre injection instabilities and reflects a different instrumental origin, not previously documented.}

\begin{figure}
    \centering
    \includegraphics[width=0.8\linewidth]{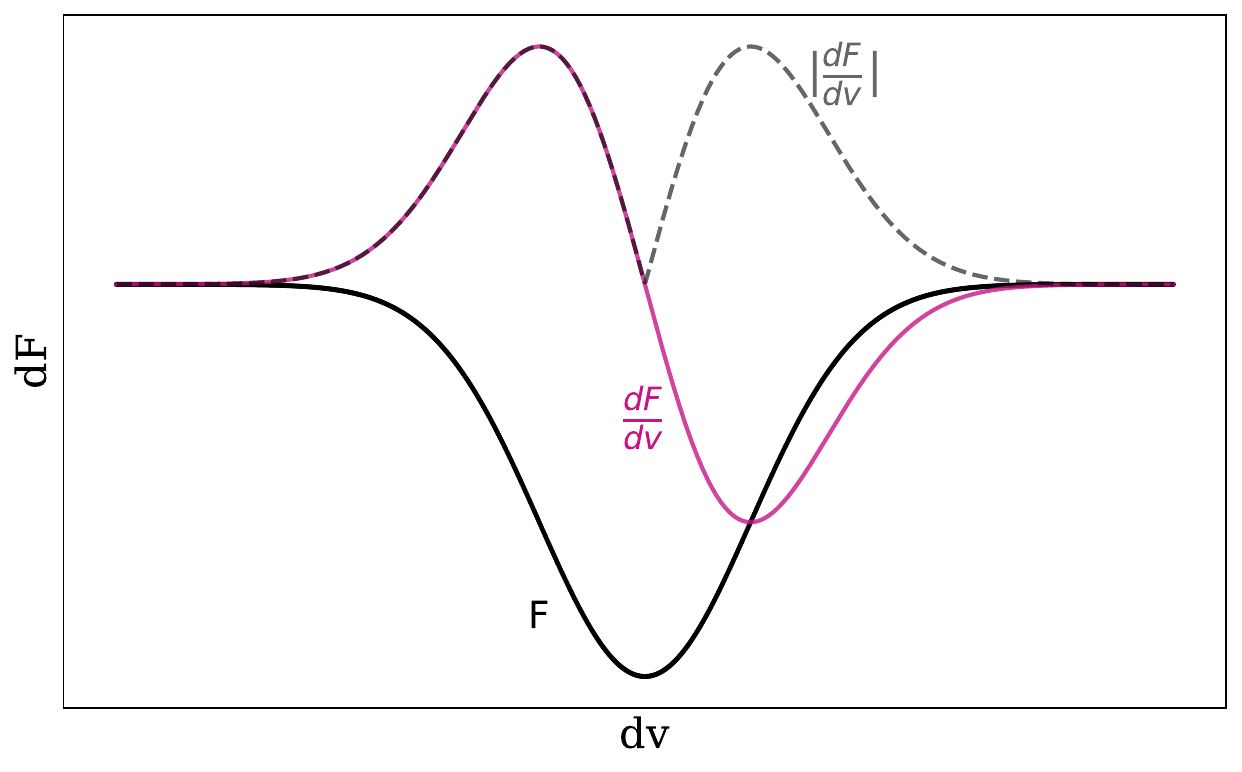}
    \caption{An illustration of a line profile and its derivative. The formation temperature is different for different parts of the profile. Averaging the formation temperature across the profile, weighted by the flux gradient with respect to velocity ( $|\frac{\rm df}{\rm dv}|$, shown dashed), yields the temperature in the layers from which most of the velocity signal originates.}
    \label{fig:CCF_deriv}
\end{figure}
To characterise relative flux variations over the area of the detector, we subdivided the extracted 2D spectrum array into a set of 4x4 contiguous blocks, creating a set of 16 "light curves" from their photon counts. We worked on the assumption that the instrumental contributions to the radial velocity could be modelled as linear combinations of these light curves.

We placed the sixteen 290-epoch light curves in an array $\bm{A} \in \mathbb{R}^{290 \times 16}$. To ensure that we used only relative (rather than absolute) flux variations in the modelling, the light curves were centred by subtracting their column means and standardised by dividing the result by the column standard deviation.  The result is illustrated in Fig.~\ref{fig:IllumArray}. The reduced singular-value decomposition of $\bm{A}$ is then
$$
\bm{A} = \bm{U} \bm{S} \bm{V}^\top,
$$
where $\bm{U} \in \mathbb{R}^{290 \times 16}$ is the left-singular matrix, $\bm{S} \in \mathbb{R}^{16 \times 16}$ is a diagonal matrix of singular values, and $\bm{V}^\top \in \mathbb{R}^{16 \times 16}$ is the right-singular matrix whose rows represent orthonormal modes of light variation across the 16 sub-regions of the CCD.

Our mitigation procedure involves projecting the radial-velocity time series into a subspace of $\bm{U}$ with reduced rank $k$:
$$
\bm{v}_{\text{proj}} = \bm{U}_k \bm{U}_k^\top \bm{v}.
$$
To minimise noise in the reconstruction, we determined the optimal rank and ordering of the columns of $\bm{U}$. We used the iterative procedure devised by \citet {ACC2021} and \citet{OuldElhkim2023}, to identify the sequence of columns that gives the fastest reduction in the statistic
$$
\chi^2 = (\bm{v} - \bm{v}_{\text{proj}})^\top \bm{\Sigma}_v^{-1} (\bm{v} - \bm{v}_{\text{proj}})
$$ 
and the Bayesian Information Criterion 
$$
\text{BIC} = \chi^2 + k \ln(n).
$$
Here $\bm{\Sigma}_v^{-1}$ is the inverse variance matrix of $\bm{v}$, $k$ is the rank of the projection, and $n$ is the number of elements in $\bm{v}$. The values of these two statistics are plotted as a function of $k$ in Fig.~\ref{fig:rearrange_U}. The BIC is minimised between $k=4$ and $k=6$. Tests with $4\leq k \leq 6$ showed little change in the pattern of residual radial velocities, indicating that the leading basis vectors capture all important details in the structure of the systematic error signal while avoiding over-fitting noise. We adopt a rank $k=4$ reconstruction, retaining only those modes of variation that produce a significant reduction in the BIC.

The reconstructed contribution to the RV from variations in CCD illumination is plotted in orange in Fig.~\ref{fig:CCD_illum_correction}, and has an RMS scatter of 48.6 cm s$^{-1}$. We conclude that a subset of the modes of CCD illumination variability recorded on both nights project strongly onto the radial velocity and that using them for linear de-trending provides effective mitigation.

Subtraction of the rank $k=4$ model from the original data brings the RMS scatter down from 70.6 cm s$^{-1}$ over the two nights to 52.9 cm s$^{-1}$ for the residual astrophysical plus noise signal plotted in green in Fig.~\ref{fig:CCD_illum_correction}. These detrended residuals were used in all subsequent analyses of the granulation signal. 

The significant dispersion observed in the EXPRES RVs (top panel of Fig. \ref{fig:DiagnosticRatios}) can be attributed to sub-optimal observing conditions, specifically poor seeing due to atmospheric smoke.. Consequently, the EXPRES RVs may not be well-suited for studying this instrumental effect. In contrast to the HARPS-N RVs, applying the same correction to the EXPRES RVs did not result in any notable improvements. Since this new instrumental signature is being identified for the first time on HARPS-N in this work, we also assess its expected contribution on the full lifetime of the instrument in Appendix~\ref{sec:LBL}.

\section{Analysis}\label{sec:analysis}

The main goal of this study is to detect and characterise the photospheric granulation signature in the spectral and RV domain. In essence, we seek to identify the stellar activity signal responsible for RV fluctuations occurring on minute-to-hour timescales.

\begin{figure}
    \centering
    \includegraphics[width=\linewidth]{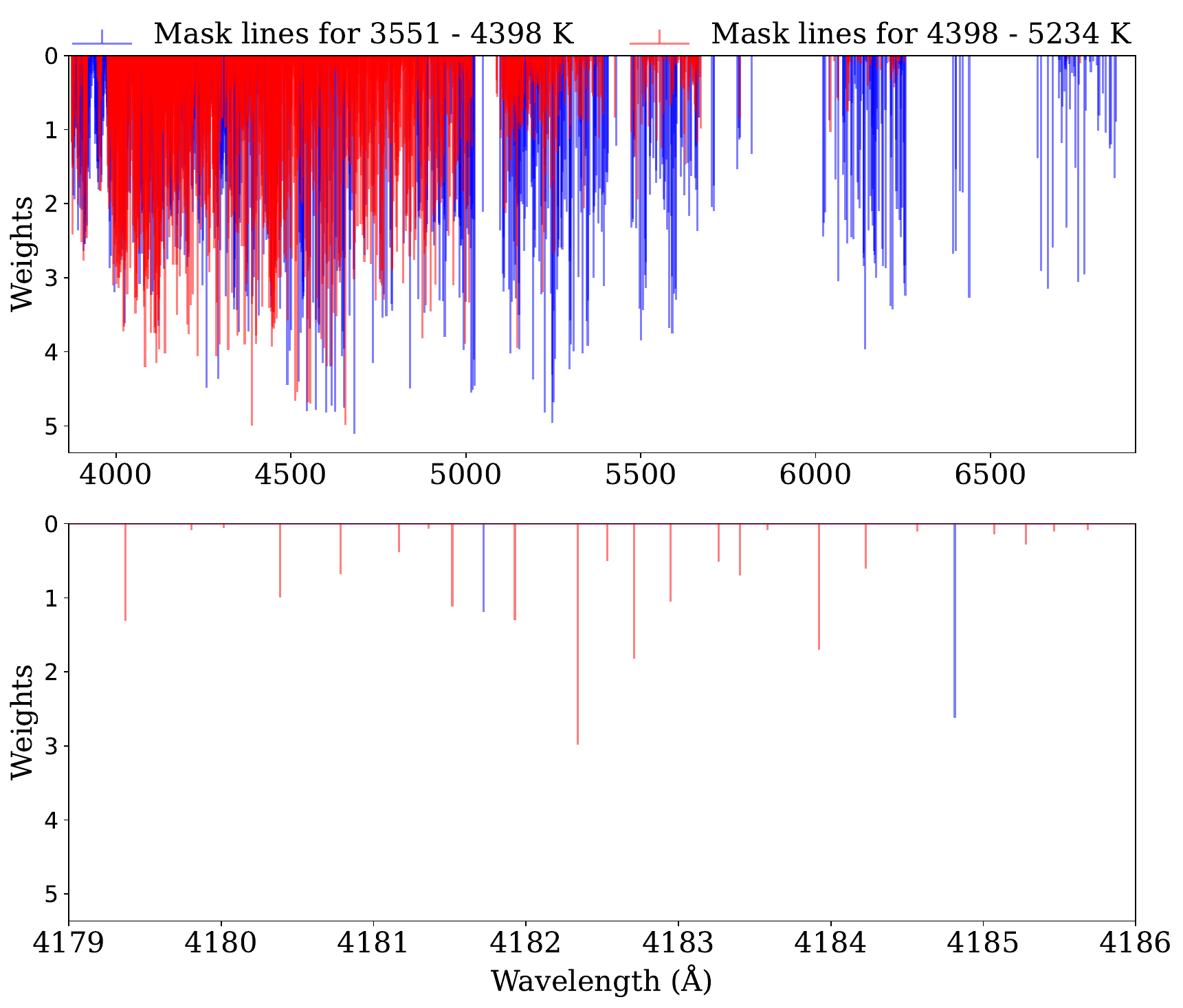}
    \caption{ \textit{Top} : The ESPRESSO K2 line mask showing distinct lines chosen based on their formation temperatures, as explained in Sect.~\ref{sec:temp_ccf_masks}. These temperature-sensitive masks are used to create different CCFs using lines formed at different photospheric depths. The y-axes represent the weights applied in the ESPRESSO DRS mask, tailored to exclude lines originating from the telluric-affected wavelength regions. \textit{Bottom}: Same zoomed-in wavelength window as in Fig. \ref{fig:spectralwindow}.}
    \label{fig: partitioned mask}
\end{figure}

\subsection{Formation temperature formalism}\label{sec:temp_formalism}
Lines of different depths probe various atmospheric layers, showing different intensity-velocity correlations within granules, thus affecting their sensitivity to convective blueshift \citep{Meunier2017}. However, line depths do not accurately indicate formation height in the photosphere. Using formation temperature as a proxy allows for better arrangement of lines by altitude in the atmosphere \citep{Ellwarth2023}.

\begin{figure*}
    \centering
    \includegraphics[width=0.36\linewidth]{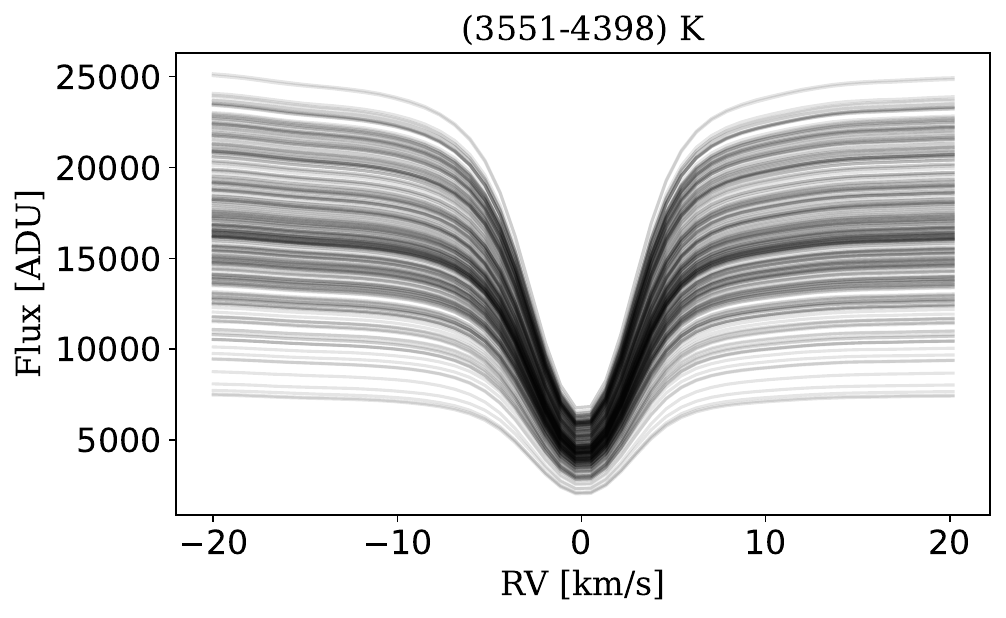}\hfill
    \includegraphics[width=0.48\linewidth]{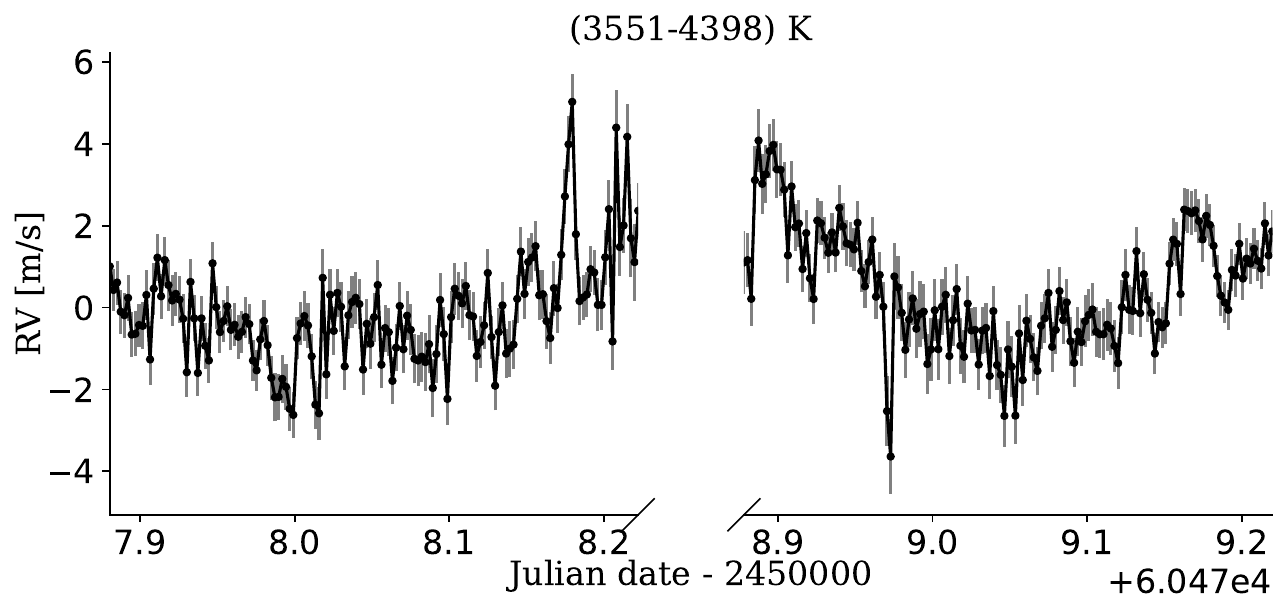}
    \includegraphics[width=0.36\linewidth]{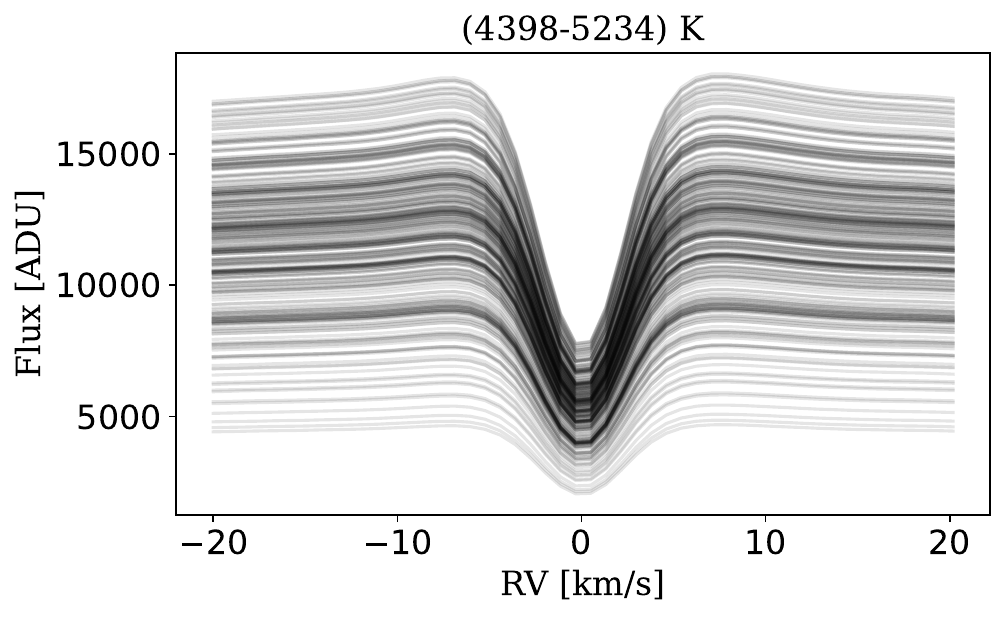}\hfill
    \includegraphics[width=0.48\linewidth]{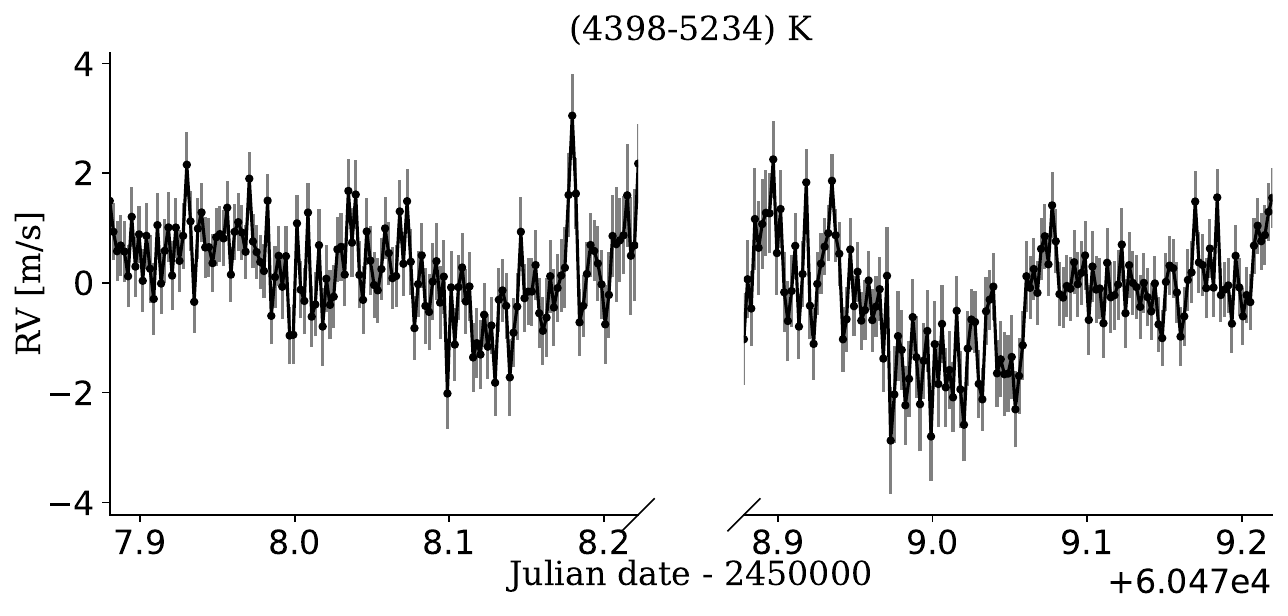}
    \caption{\textit{Left}: Each set of 290 CCFs was computed from the HARPS-N spectra of HD\,166620 obtained over two nights. These CCFs were computed using subsets of the ESPRESSO K2 line mask, partitioned into two $T_{1/2}$ ranges: [3551, 4398] K and [4398, 5234] K, as shown in Fig.~\ref{fig: partitioned mask}. Since ARVE computes CCFs in the same units as the input spectra, the resulting CCFs display the actual average flux counts. The observed flux offsets likely reflect variations caused by changes in airmass. \textit{Right}: The RV time series extracted from the temperature-partitioned CCFs in the left panels. Note: These are not corrected for the instrumental issue discussed in Section \ref{sec:instrumental}.}
    \label{fig:hotcoolCCFs}
\end{figure*}

Employing the methodology outlined by \citet{AlMoulla2022} and \citet{AlMoulla2024}, we assess the relative susceptibility to granulation of line segments formed at different photospheric temperature ranges. Specifically, we investigate whether RV perturbations are more pronounced in spectral intervals formed deeper in the photosphere (where the contrasts in radial flow velocities and temperatures are stronger) or in the shallower layers.

The RVs at different temperature ranges are extracted with the Analysing Radial Velocity Elements\footnote{\url{https://github.com/almoulla/arve}} (\texttt{ARVE}; Al Moulla, submitted) code, which makes use of synthetic spectra from \texttt{PySME}\footnote{\url{https://github.com/AWehrhahn/SME}} \citep{Wehrhahn2023}. \textcolor{black}{Specifically, the spectra are synthesized with one-dimensional radiative transfer at local thermodynamic equilibrium, using MARCS \citep{Gustafsson2008} model atmospheres and VALD3\footnote{\url{https://vald.astro.uu.se}} \citep{Ryabchikova2015} line lists. From the syntheses, the flux contribution functions are extracted to determine} the photospheric temperature below which 50\% of the emergent flux is formed at any given sampled wavelength, referred to as the average formation temperature and denoted as $T_{1/2}$. \texttt{ARVE} has a library of precomputed spectra for different spectral types at visible and near-infrared wavelengths; by providing a target name and a reference spectrum, the relevant synthetic \textcolor{black}{spectrum is} fetched and resampled to the wavelength grid of the reference spectrum. Fig.~\ref{fig:spectralwindow} shows the synthetic flux spectrum for spectral type K2, \textcolor{black}{colour-coded by its $T_{1/2}$ values.}

\subsubsection{Constructing temperature-sensitive CCF masks - \textcolor{black}{$ {\rm CCF}_{T}$} method}\label{sec:temp_ccf_masks}

In \citet{AlMoulla2022}, temperature-dependent RVs were extracted by template-matching spectral segments formed at a certain temperature range against a reference spectrum built from all observed spectra of the star, and thereafter averaging together the RVs of all flux segments formed within the same temperature range. This is essentially the same as the LBL method but generalised to consider parts of spectral lines instead of the entire lines. Hereafter, we will refer to this extraction method as \textcolor{black}{$ {\rm LBL}_{T}$}, in the context of a temperature range. For this study, we aim to compute temperature-dependent RVs using the \textcolor{black}{$ {\rm CCF}_{T}$} method. The motivation for this is to (1) minimise any biases inherent to a single extraction method by employing two RV extraction techniques in parallel, and (2) to 
construct temperature-dependent CCFs which could thereafter be fed to activity-mitigation techniques specifically relying on the CCF format, such as \textsc{scalpels} \citep{ACC2021} or \textit{CALM} \citep{deBeurs2024}.

The temperature-sensitive CCF line masks are constructed as follows. We begin with the spectral lines listed in the ESPRESSO K2 mask, and cross-match them with the line list from \texttt{ARVE} to retrieve their estimated wavelength bounds. For each line, the weight $w$ associated with the wavelength at pixel $i$ can be represented as
\begin{equation}
    w_{i} = \left| \frac{\mathrm{d}F}{\mathrm{d}v} \right|_{i} \,,
\end{equation}
where $F$ is the flux and $v$ the velocity shift away from the line centre. We can then define the flux-gradient-weighted temperature for each line as
\begin{equation}
    \hat{T}_{1/2} = \frac{\sum w_{i} T_{1/2,i}}{\sum w_{i}} \,,
\end{equation}
where the summation is performed over all pixels within the bounds of the line. This will ensure putting higher weights on the wings of the line, which contains more RV information than the core (see Fig.~\ref{fig:CCF_deriv}).

To compute the CCF for a specific temperature range, the calculation includes only the mask lines with $\hat{T}_{1/2}$ values within that selected range. Fig.~\ref{fig: partitioned mask} shows a temperature partition made on the ESPRESSO K2 mask, and Fig.~\ref{fig:hotcoolCCFs} shows the associated CCFs and RVs for the HARPS-N observations of HD\,166620.

We remark that these temperature-dependent CCF RVs are fundamentally different from their \textcolor{black}{$ {\rm LBL}_{T}$} counterparts. While the \textcolor{black}{$ {\rm LBL}_{T}$} method only considers spectral segments (segmented in flux) formed at a certain temperature range, the \textcolor{black}{$ {\rm CCF}_{T}$} method still includes entire lines, parts of which might be formed outside the temperature range. The lines used in the \textcolor{black}{$ {\rm CCF}_{T}$} calculation should, however, still be weighted toward lines mostly formed in the temperature range, and should thus produce similar results. To find out whether this holds is another motivation for using the two RV extraction methods in parallel.


\subsubsection{Selecting formation temperature ranges}\label{sec:temp_ranges}

To investigate the sensitivity of spectral regions formed at different photospheric depths, to variations induced by granulation, we measure the RV as a function of the average formation temperature, $T_{1/2}$. Increasing $T_{1/2}$ implies deeper layers in
the photosphere. We have no \textit{a priori} knowledge about which temperature range yields the quietest RV time series suitable for granulation characterisation. Therefore, we perform an empirical search, varying the range within the values that the average formation temperature takes on for this specific spectral type and wavelength range. We select two distinct temperature ranges: [3551, 4398]\,K and [4398, 5234]\,K, ensuring comparable RV content and variance. The CCFs and RVs extracted from these ranges, representing the cooler upper layers and the hotter deeper layers of the photosphere, are displayed in the upper and lower panels of Fig. \ref{fig:hotcoolCCFs}, respectively. We also computed the \textcolor{black}{$ {\rm LBL}_{T}$} RVs for these temperature ranges.

To account for the instrumental systematics discussed in Sect.~\ref{sec:instrumental}, the RV time series extracted using the \textcolor{black}{$ {\rm CCF}_{T}$} methods underwent CCD-illumination decorrelation. The CCD illumination correction is individually performed for each of the 2x2 RV time series using unique sets of $\bm{U}$ (basis) vectors. The resulting \textcolor{black}{$ {\rm CCF}_{T}$} and \textcolor{black}{$ {\rm LBL}_{T}$} RVs are shown in Fig.~\ref{fig:28CCFRVs}.

\section{Results}
\label{sec:results}

\subsection{Structure function analysis} \label{sec:structure_function}

\subsubsection{Method}
The structure function (SF) is a tool used to investigate variability within a time series ($f$) as a function of timescale ($\tau$). For continuous data the SF is defined as
\begin{equation}\label{eq:SF_cont}
    {\rm SF}(\tau) = \langle(f(t)-f(t+\tau))^2\rangle
\end{equation}
where $\tau$ is the time separation between two data points in a pair (i.e., $\tau = |t_1-t_2|$) and the angular brackets refer to the fact that the SF is averaged over all data points with time separation $\tau$. For discrete data, the SF is normally calculated in logarithmic bins of $\tau$ as
\begin{equation}
    {\rm SF}(\tau_1,\tau_2) = \frac{1}{N_{\rm p}} \sum_{i,j}(f_i-f_j)^2
\label{SF_def2}
\end{equation}
where $N_{\rm p}$ is the number of pairs in the bin, $\tau_1$ and $\tau_2$ are the bounds of the timescale bin such that $|t_i-t_j|$ in the bin are between these bounds, where $t_i$ and $t_j$ are the times at $f_i$ and $f_j$ respectively. In the cases investigated here, $f_i$ and $f_j$ refer to RV data.
For more detailed discussions of SFs, see for example: 
\citet{1985ApJ...296...46S,2010MNRAS.404..931E,2020MNRAS.491.5035S,2022MNRAS.514.2736L}.
In the limit of a continuous uncorrelated signal, ${\rm SF}(\tau) \equiv 2 \sigma^2$, where $\sigma$ is the root-mean-square (RMS) scatter of the signal \citep[see][]{Lakeland2024} .
We therefore plot $\sqrt{\nicefrac{1}{2}\,{\rm SF}}$ in our analysis.
\begin{figure}
    \centering
    \includegraphics[width=\linewidth]{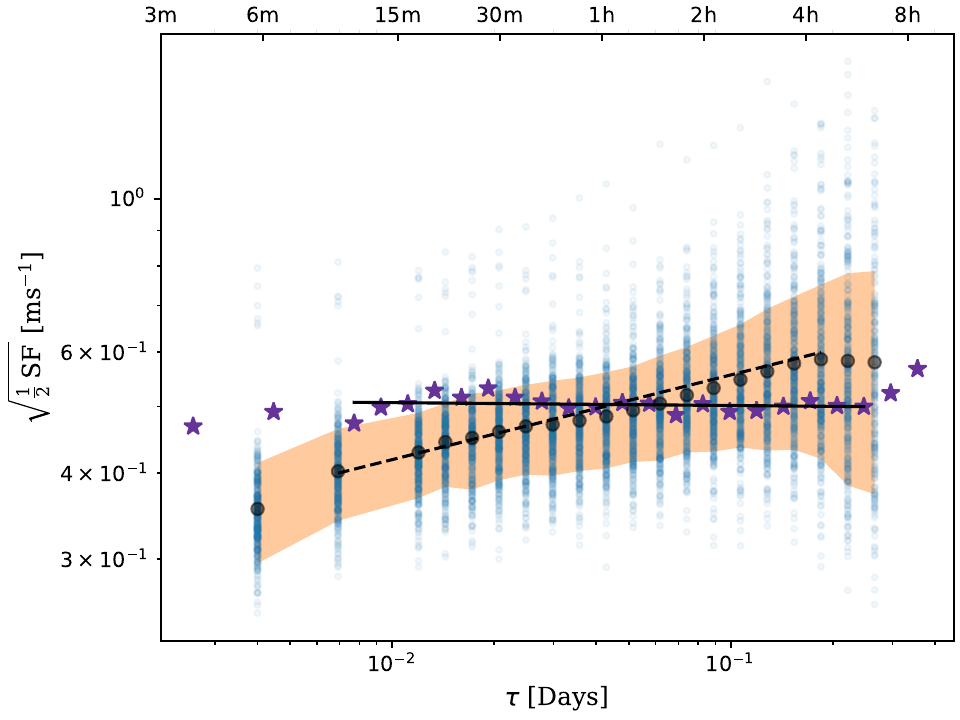}
    \caption{$\sqrt{\nicefrac{1}{2}\,{\rm SF}}$ (SF: structure function), which is equivalent to RMS scatter, of HD\,166620 (purple stars) compared with $\sqrt{\nicefrac{1}{2}\,{\rm SF}}$ of the Sun in many 2 day chunks (blue circles) as a function of timescale. The plot also shows the mean of the solar SFs (black circles) and a one standard deviation shaded region in orange. The solid line and the dashed line are based on the fits to the SF ($\log{\rm SF} = m\,\log{\tau} + k$) of the HD\,166620 data and the mean SF of the solar SFs, plotted is the square root of half of the linear fit to the SF.}
    \label{fig:SF_sun_HD166620}
\end{figure}
\subsubsection{\textcolor{black}{Illumination-corrected radial velocities}}\label{sec:SF_ICRV}


To compare the signals present in the RV time series of HD~166620 to the levels of granulation and supergranulation present in the Sun \citep[see e.g.,][]{AlMoulla2023, Lakeland2024}, we take \textcolor{black}{around seven} years of public HARPS-N solar data \citep[][\color{black}Dumusque et al. in prep.]{ACC2019, Dumusque2021} and divide the time series into successive chunks of two days to best emulate the HD~166620 observation strategy. 

If the p-modes in the Sun are not sufficiently averaged out they quickly overwhelm other signals in the SF due to its cumulative-like nature\textcolor{black}{.
It is also important to ensure that the p-mode reduction is similar between HD\,166620 and the Sun for a fair comparison,} hence, the \textcolor{black}{solar} data has a cadence of \textcolor{black}{5}\,minutes ensuring the p-modes are averaged out \textcolor{black}{but binned for approximately one cycle like HD\,166620} \citep{Dumusque2011,Chaplin2019}.
We note that the timescales that are available for the Sun when looking at chunks of two days are more limited than that \textcolor{black}{of} HD\,166620 due to the length of the observation window in a day or night.

The number of pairs of data points in a SF bin ($N_{\rm P}$ in Equation \ref{SF_def2})
is important to consider as having few pairs can lead to noisy and unreliable SFs. By examining the solar SFs, we determine that a minimum of 30 RV point pairs per SF bin for HD~166620 effectively reduces noise in the SFs without misrepresenting the shape and behaviour of the data. This selection of a minimum number of pairs does not cause any SF bins for HD\,166620 to be removed. 
In addition, we only considered structure function timescales less than $\sim$ 9.3 hours. 
This is to avoid comparing observations which bridge the diurnal gap in the data, and removes the potential effect of any remnant inter-night instrumental effects.

Fig.~\ref{fig:SF_sun_HD166620} shows a comparison between the SFs of \textcolor{black}{the illumination-corrected RVs of} HD\,166620 and the Sun. The SF of the HD\,166620 RV data is much flatter than the SFs we calculate from solar RV data at this timescale. HD\,166620 may show a hump feature at around $\tau=0.016\,{\rm days}\,(23\,{\rm minutes})$ which could be indicative of a component of correlated noise or periodic signal even though the SF is so flat.

\begin{figure}
    \centering
    \includegraphics[width=\linewidth]{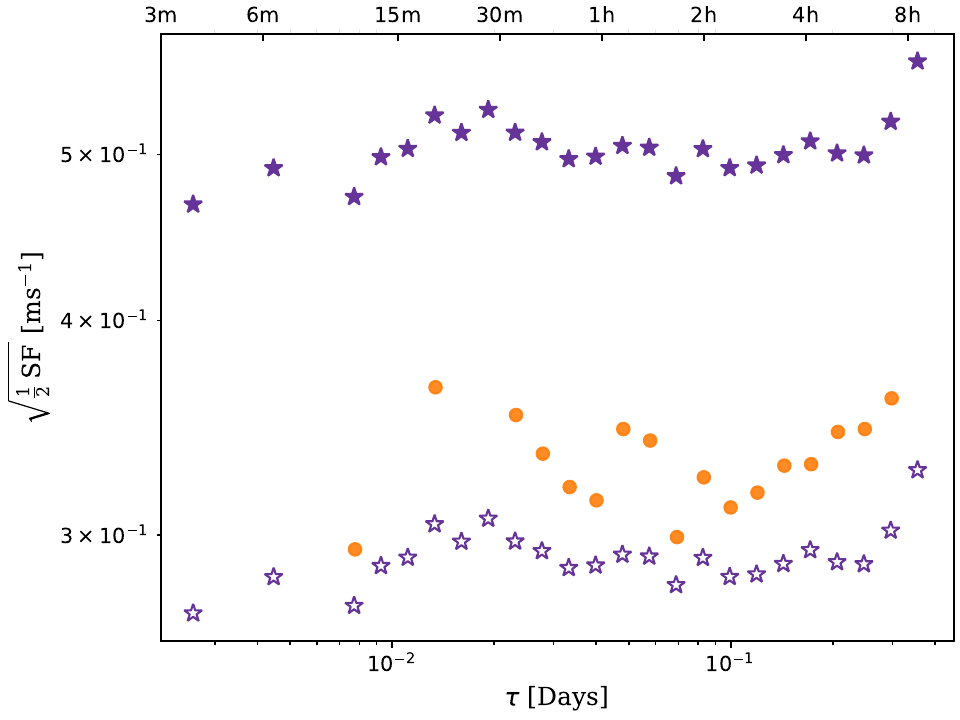}
    \caption{$\sqrt{\nicefrac{1}{2}\,{\rm SF}}$ of HD\,166620 as a function of timescale. The dark purple stars give the HD\,166620 RVs as processed in Section \ref{sec:obs} and the purple \textcolor{black}{unfilled stars} give the same SF divided by $\sqrt{3}$. The orange circles give the SF of the HD\,166620 RVs binned in groups of 3. The fall in the SF of HD\,166620 is not quite $\sqrt{3}$ suggesting that the RVs are not completely consistent with uncorrelated noise despite the flatness.}
    \label{fig:SF_HD_3s}
\end{figure}

To quantify how flat the SF of HD\,166620 is compared to the Sun, we used a least squares method to fit a linear relation to the SFs of HD166620. We fit to the timescales around 10\,minutes to 6\,hours in log space using $\log{\rm SF} = m\,\log{\tau} + k$ (note this is fit to the SF and not $\sqrt{\nicefrac{1}{2}\,{\rm SF}}$). For both the Sun and HD 166620, the uncertainties for each SF value are set to 1 and scaled such that the reduced $\chi^2$ is unity\footnote{See the Appendix of \citet{2020MNRAS.491.5035S} for a discussion of the difficulties in estimating uncertainties in structure functions.}.
We find the gradient of the solar SF to be \textcolor{black}{$0.246\pm0.009$} and HD\,166620 to be $-0.008 \pm 0.011$ which is consistent with zero. Hence, the RV time series of HD~166620 is likely consistent with uncorrelated noise at these timescales \citep{2020MNRAS.491.5035S}.
This could be due to a genuinely uncorrelated signal such as photon noise, or a signal such as granulation which is correlated at short timescales, but we would expect to be uncorrelated on these timescales. 
This contrasts with the Sun is dominated by correlated noise on similar timescales, probably caused by a supergranulation component \citep{Lakeland2024}.
\textcolor{black}{We also experimented with the solar data binned to longer exposure times and this causes the gradient of the structure function to increase, showing that the choice of exposure time (or effective exposure time) is important to consider before making comparisons between HD\,166620 and the Sun.}

With SFs as a function of $\tau$, correlated noise appears as a slope and periodic signals tend to be shown as a rise and then fall (hump) \citep[e.g.][]{2020MNRAS.491.5035S,2022MNRAS.514.2736L}.
The hump feature in the SF of HD\,166620 around the $23\,{\rm minute}$ timescale
could be consistent with a component of correlated noise or some periodic signal. As $\sqrt{\nicefrac{1}{2}\,{\rm SF}}$ is equivalent to RMS scatter for uncorrelated noise \citep{Lakeland2024}, we can test this by binning the RV data of HD\,166620.
If the SF at these timescales were consistent with only uncorrelated noise, we would expect $\sqrt{\nicefrac{1}{2}\,{\rm SF}}$ to fall by $\sqrt{n}$, where $n$ is the number of RV data points binned together.
If $\sqrt{\nicefrac{1}{2}\,{\rm SF}}$ does not fall as we expect for uncorrelated noise, this would hint that there is some correlation within our data set.
We choose to bin the RV time series in groups of three (ensuring that observations either side of the daytime gap are not binned together) as a majority of the SF bins contain enough pairs of data points while still probing the hump feature at $23\,{\rm minutes}$. The bin at $\tau \approx 8.6\,{\rm hours}\,(0.36\,{\rm days})$ is not plotted as it only contains six pairs of data points.
Fig.~\ref{fig:SF_HD_3s} shows that $\sqrt{\nicefrac{1}{2}\,{\rm SF}}$ instead falls as $\sim\sqrt{2.32}$ instead of $\sqrt{3}$, showing there is some correlated noise in the HD\,166620 RVs.

To estimate the correlated noise amplitude of HD\,166620 we assume following model for the RV data. If the HD\,166620 data has RMS scatter $\sigma_1$ which is a combination of correlated noise with amplitude $\sigma_{\rm c}$ and uncorrelated noise with an amplitude of $\sigma_{\rm u}$ then,
\begin{equation}\label{eq:toy_model_RV1s}
    {\sigma_1}^2 = {\sigma_{\rm u}}^2 + {\sigma_{\rm c}}^2.
\end{equation}
We assume that the correlated has a timescale longer than the binned RV time series but significantly shorter than one night of data.
When we bin the RV data in groups of three, resulting in the RMS scatter $\sigma_3$, the uncorrelated noise amplitude is reduced by $\sqrt{3}$ but we assume that the uncorrelated noise is unaffected hence,
\begin{equation}
    {\sigma_3}^2 = \frac{{\sigma_{\rm u}}^2}{3} + {\sigma_{\rm c}}^2.
\end{equation}
Combining this with the above statement that we observe the SF of the binned-in-threes RVs to fall by $\sqrt{2.32}$, we obtain
\begin{equation}\label{eq:toy_model_RV3s}
    \frac{3}{2.32} {\sigma_1}^2 = {\sigma_{\rm u}}^2 + 3{\sigma_{\rm c}}^2.
\end{equation}
Equations~\eqref{eq:toy_model_RV1s} and~\eqref{eq:toy_model_RV3s}
result in an estimate of the amplitude of the uncorrelated noise of
\begin{equation}\label{eq:corr_noise}
    \sigma_{\rm c} = \sqrt{\frac{3/2.32-1}{2}}\,\sigma_1 \simeq0.38\,\sigma_1.
\end{equation}
We calculate the mean of $\sqrt{\nicefrac{1}{2}\,{\rm SF}}$ of HD\,166620 at the available timescales to be 50\cms. The uncertainty in $\sigma_{\rm c}$ is dominated by the uncertainty in $\sigma_3$ which we assume to be the standard deviation of its scatter.
Hence we determine the \textcolor{black}{correlated} noise amplitude to be $19\pm5\,$\cms.
 In summary, the structure function indicates there is correlated noise on timescales $\tau \simeq 20\,$minutes with an amplitude of $\simeq20\,$\cms.

We do not see compelling evidence of supergranulation on the timescales and the amplitudes seen in the Sun.
At short timescales there is largely uncorrelated noise at amplitudes exceeding that of supergranulation in the Sun (Fig~\ref{fig:SF_sun_HD166620}).
If there were solar like supergranulation, at longer timescales we would expect the SF to begin rising at around \textcolor{black}{0.05}~days (where the SF of HD\,166620 crosses the SF of the Sun). In fact, it remains flat, but also lies within one standard deviation of many solar SFs. Hence, more data are required to detect or rule out supergranulation at solar levels.

\subsubsection{\textcolor{black}{Temperature dependent radial velocities}}

\begin{figure}
    \centering
    \includegraphics[width=\linewidth]{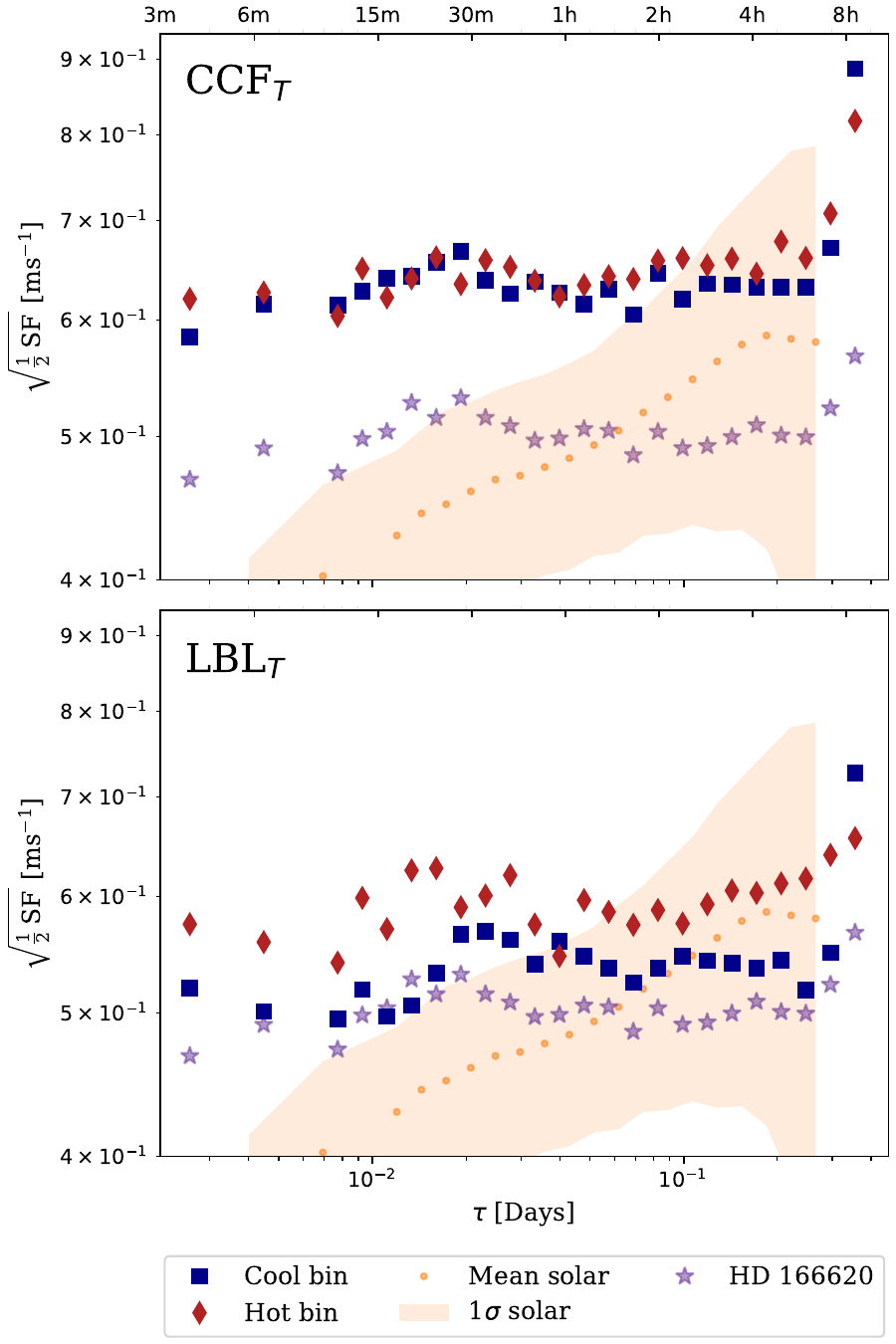}
    \caption{\textcolor{black}{$\sqrt{\nicefrac{1}{2}\,{\rm SF}}$ versus timescale for the temperature binned radial velocities. \textbf{Top}: CCF$_T$. \textbf{Bottom}: LBL$_T$. Red diamonds give the hotter RV bin and blue squares give the cooler RV bin. The structure function of the illumination corrected RVs and the mean solar structure function are given as purple stars and orange circles respectively (with one standard deviation of the solar RVs given as an orange shaded area) to use as reference points.}}
    \label{fig:SF_temp_bins}
\end{figure}

\textcolor{black}{
We calculate SFs of the ${\rm CCF}_{T}$ and ${\rm LBL}_{T}$ temperature binned RVs which are presented in Fig.~\ref{fig:SF_temp_bins}.
We see that the RMS as a function of timescale has generally increased in comparison to the illumination corrected RVs which is reasonable as the spectra have been split into two datasets, a hot bin and a cool bin.
We note that the variability in the time series between the cool and hot bins is more similar in the ${\rm CCF}_{T}$ method than the ${\rm LBL}_{T}$.
This is likely a result of the fundamental differences between the techniques where temperature bin widths were selected in the ${\rm CCF}_{T}$ method but not in the ${\rm LBL}_{T}$ method.
This does mean that the SF analysis of the ${\rm CCF}_{T}$ agrees that the temperature ranges selected in Section~\ref{sec:temp_ranges} are such that the variability in the hot and cold bins are comparable, even as a function of timescale.}

\begin{table}
\label{tab:SF_temp_bins}
\centering
\caption{Correlated and uncorrelated signal amplitudes from the temperature dependant radial velocities described in Section~\ref{sec:analysis} based on the structure function analysis with Equation~\ref{eq:toy_model_general_RV3s}.}
\begin{tabular}{cccc}
\hline
Method          & Signal        & Cool bin amplitude & Hot bin amplitude\\
                &               & [\ms]              &[\ms]          \\\hline
${\rm CCF}_{T}$ & correlated    & $0.22\pm0.02$      & $0.10\pm0.03$ \\
                & uncorrelated  & $0.59\pm0.02$      & $0.63\pm0.02$ \\\hline
${\rm LBL}_{T}$ & correlated    & $0.22\pm0.03$      & $0.18\pm0.03$ \\
                & uncorrelated  & $0.49\pm0.02$      & $0.56\pm0.02$ \\
\hline 
\end{tabular}
\end{table}

\textcolor{black}{
We repeat the binning-in-threes experiment described in Section~\ref{sec:SF_ICRV}.
A more general form of Equation~\ref{eq:toy_model_RV3s} is given by
\begin{equation}\label{eq:toy_model_general_RV3s}
    \frac{3}{{\Delta_{\rm f}}^2} {\sigma_1}^2 = {\sigma_{\rm u}}^2 + 3{\sigma_{\rm c}}^2.
\end{equation}
Where $\Delta_{\rm f} = \sigma_1/\sigma_3$ with $\sigma_1$ and $\sigma_3$ calculated as the average over $\tau$ of $\sqrt{\nicefrac{1}{2}\,{\rm SF}}$.
The extracted uncertainties of the uncorrelated and correlated noise are dominated by the uncertainty in $\sigma_1$ and $\sigma_3$ respectively allowing us to determine their amplitudes as presented in Table~\ref{tab:SF_temp_bins}.
We find that the hot bin of ${\rm CCF}_{T}$ contains the least correlated signal which is further shown in Figure~\ref{fig:SF_CCF_threes}.
Both bins of the ${\rm LBL}_{T}$ method behave similarly to the ${\rm CCF}_{T}$ cool bins when the RVs are binned in time groups of three data points, so other than the hot bin ${\rm CCF}_{T}$ RVs, the binned-in-threes RVs always sit above the predicted fall for a time series of only uncorrelated noise ($\sqrt{\nicefrac{1}{2}\,{\rm SF}}/\sqrt{3}$).
For the hot bin ${\rm CCF}_{T}$ RVs however, four out of the eleven data points used to calculate the average SF even sit below $\sqrt{\nicefrac{1}{2}\,{\rm SF}}/\sqrt{3}$, lending more weight to the statement that the hot bin ${\rm CCF}_{T}$ RVs contain less correlated signal.
We thus recommend when looking at the ${\rm CCF}_{T}$ RVs, to characterise the correlated signal, one should use the cooler bin RVs.
}

\subsection{GP analysis \& Power spectra}

\subsubsection{Method}

Following the method of \cite{OSULLIVAN2024}, we fit a one-component Gaussian Process (GP) model to the data.  We briefly describe the method below; interested readers are referred to \citet{OSULLIVAN2024} for a more detailed explanation.

We use an over-damped harmonic oscillator implemented in the {\tt celerite2} package \citep{celerite1,celerite2} as a Simple Harmonic Oscillator (SHO) kernel with a quality factor $Q=1/\sqrt{2}$. This component is used to model the granulation signal in HD\,166620. We experimented with a two-component model, to also model the supergranulation component of the data, however, due to the limited amount of data and the expected time scale of supergranulation (around 1 day), we were unable to fit the supergranulation. Our model also includes a white noise term added to the diagonal of the covariance matrix to account for the uncertainty due to photon noise and instrumental systematics.
Overall, the GP kernel we use to model the RVs is then
\begin{equation}
    \label{eq:quiet RV Kernel}
    k_{\rm quiet} = k_{\rm gran}  + k_{\rm w},
\end{equation}
where $k_{\rm gran}$ is an aperiodic SHO kernel:
\textcolor{black}{
\begin{equation}
    \label{eq:k_SHO}
    k_{{\rm SHO},Q=\frac{1}{\sqrt{2}}} (\tau) = S_G \, \omega_G \, e^{-\frac{1}{\sqrt{2}} \omega_G \tau} \, \cos \left( \frac{\omega_G \tau}{\sqrt{2}} - \frac{\pi}{4} \right),
\end{equation}
}
with $\tau \equiv |t_i-t_j|$, representing the granulation component, and $k_{\rm w}$ is a white noise term:
\begin{equation}
    \label{eq:k_w}
    k_{\rm w} (t_i,t_j) = \delta_{ij} (\sigma_{\rm n}^2 + \sigma_{i}^2),
\end{equation}
where $\sigma_n$ is the white noise standard deviation and $\sigma_{i}$is the formal uncertainty of measurement i.

\begin{figure}
    \centering
    \includegraphics[width=0.48\textwidth]{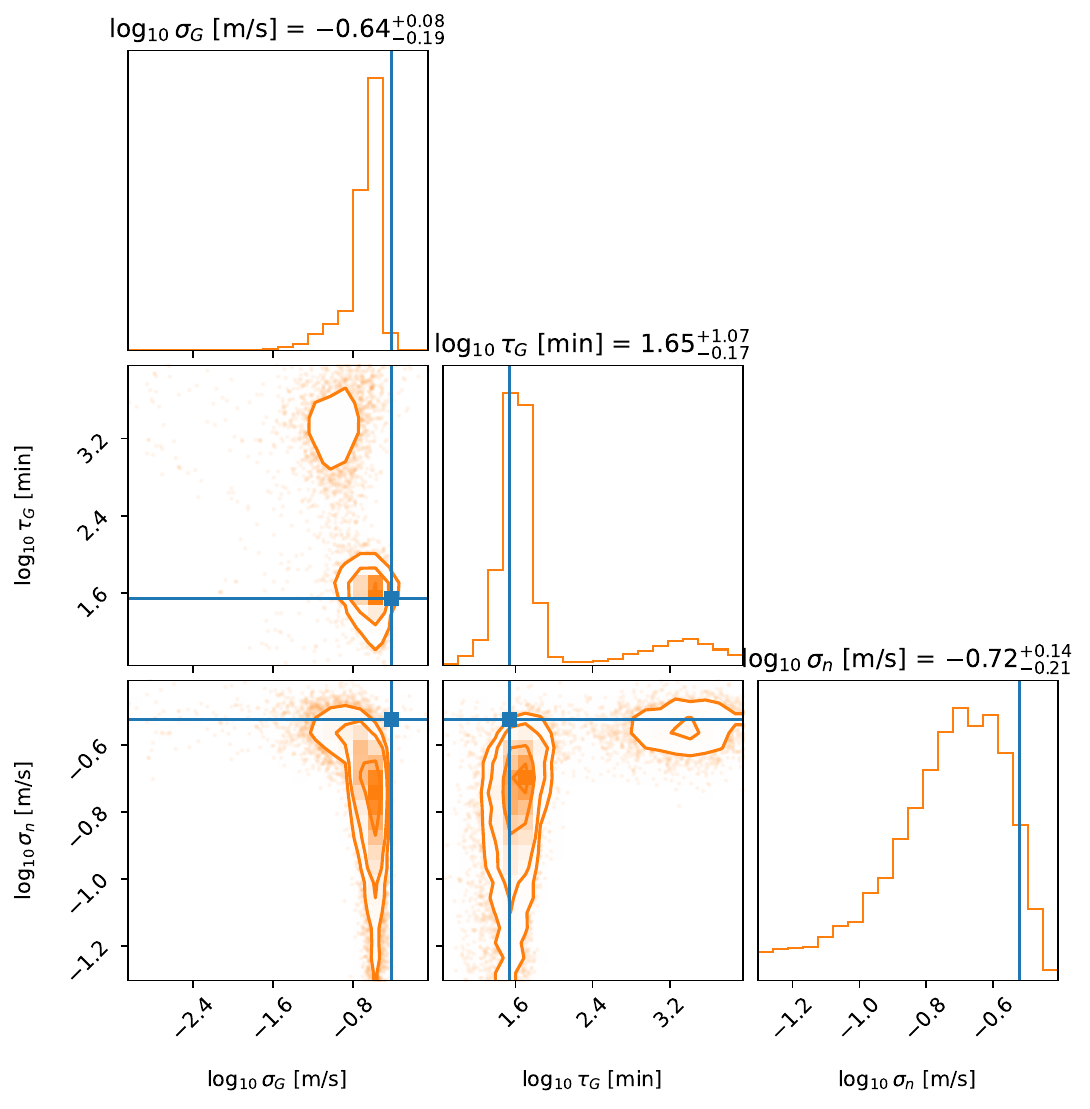}
    \caption{MCMC posterior distribution plots for the one-aperiodic GP model of the de-trended HARPS-N \textcolor{black}{DRS} RVs. \textcolor{black}{ The parameters shown are the granulation standard deviation, $ \log_{10}\sigma_G$, the granulation timescale, $\log_{10}\tau_G$, and the white noise, $\log_{10}\sigma_n$. }The 1-D posterior distributions for each parameter, marginalised over all the other parameters, are shown by the histograms in the diagonal panels. The predicted values are indicated by the blue lines. The smaller peak in the distributions likely comes from the window function. The 2-D posteriors are shown in the off-diagonal panels. }
    \label{fig:both night corner plot}
\end{figure}

Following \citet{OSULLIVAN2024}, we use a Markov Chain Monte Carlo (MCMC) to sample the joint posterior probability distribution over the three free parameters of our model, which are the characteristic amplitude, $S_G$, and angular frequency, $\omega_G$, for the granulation kernel, and the white noise standard deviation $\sigma_{\rm n}$. We used version 3 of the {\sc emcee} package \citep{2013PASP..125..306F,2019JOSS....4.1864F} to perform the MCMC sampling. The walkers are initialised in a tight Gaussian ball (with standard deviation 0.01\,dex) around a local optimum found using the {\sc minimize} function in {\sc scipy}'s {\sc optimize} module. Log$_{10}$-uniform priors are used for all the parameters, within the interval $[-10;0]$ $(m s^{-1})^2$ for $\ln S_G$, $[0;10]$ {$days^{-1}$ for $\ln \omega_G$, and $[-2;2]$ $m s^{-1}$ for the standard deviation of the white noise$\ln \sigma_{\rm n}$. The number of walkers is set to 12 (4 times the number of parameters), and the MCMC chains are run for 10\,000 steps.

\begin{figure}
    \centering
     \includegraphics[width=1\linewidth]{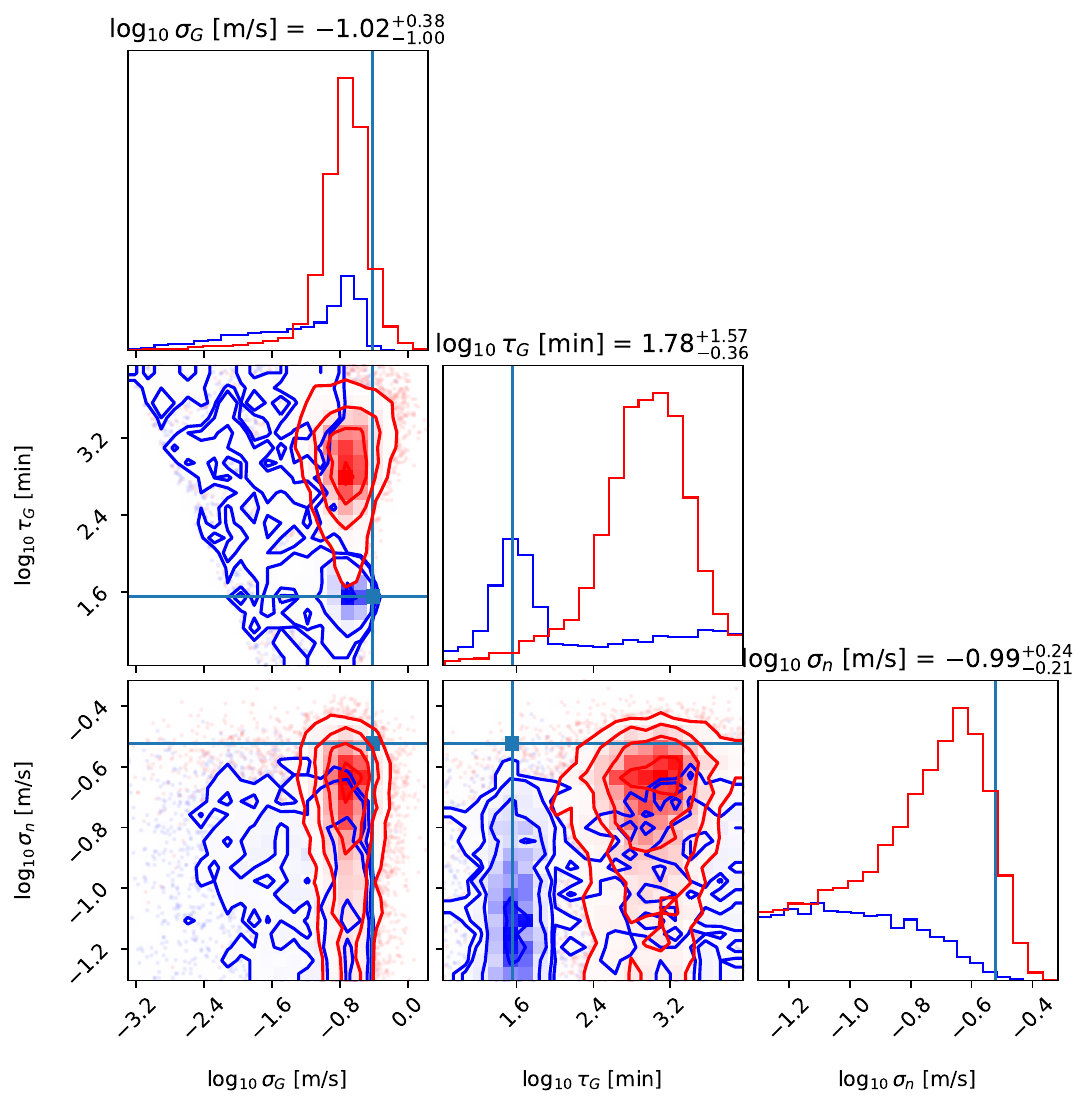}\\
     \includegraphics[width=1\linewidth]{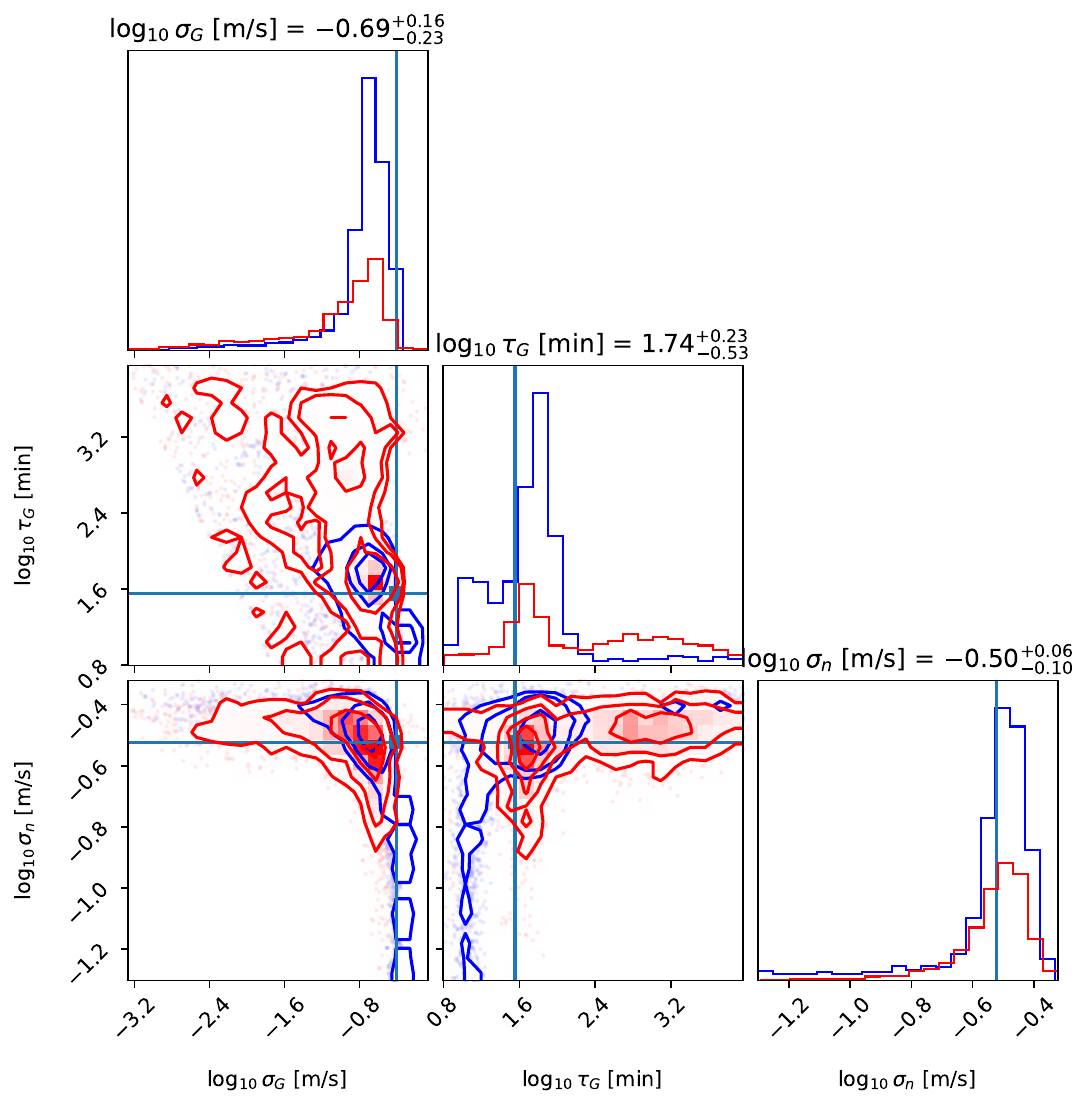}
    \caption{\textcolor{black}{  MCMC posterior distribution plots for the one-aperiodic GP model of the temperature dependent RVs.  The top column shows the results for the  $ {\rm CCF}_{T}$ RVs derived using the two temperature-partitioned CCFs. The bottom column shows the results for the $ {\rm LBL}_{T}$ RVs. The parameters shown are the granulation standard deviation, $ \sigma_G$ , the granulation timescale, $\tau_G$ , and the white noise, $\sigma_n$. The 1-D posterior distributions for each parameter, marginalised over all the other parameters, are shown by the histograms in the diagonal panels. The blue contours represent the cooler RVs, while the red contours show the results for the hotter RVs originating from the deeper layers. The titles show the best fit values for the cooler RVs. The predicted values are indicated by the blue lines. The smaller peak in the distributions likely comes from the window function. The 2-D posteriors are shown in the off-diagonal panels. }}
    \label{fig:CCF_LBL GP results}
\end{figure}

To assess convergence and select the appropriate burn-in and thinning factor, we use {\sc emcee}'s built-in functionality to estimate the auto-correlation length of the chains. Provided that the longest auto-correlation length estimate across all parameters is $\tau_{\rm max} \le 200$ steps, we consider the auto-correlation time estimates to be reliable and the chains well-converged. We then discard the first $3 \tau_{\rm max}$ samples as part of the burn-in phase, and thin the remainder of the chains by a factor $\tau_{\rm max}/4$.

\subsubsection{Results}
To make our results more accessible, we convert $\ln S_G$ and $\ln \omega_G$ to the standard deviation, $\log_{10} \sigma_G$, and the undamped period,  $\log_{10} \tau_G $.The conversions are given by the following formulas \citep{OSULLIVAN2024};
\begin{equation}
    \label{eq:variance}
    \sigma_G^2 = S_G \omega_G Q = \frac{1}{\sqrt{2}} S_G \omega_G,
\end{equation}
\begin{equation}
    \label{eq: tau}
    \tau = 2\pi/\omega_G. 
\end{equation}
We highlight that the definition of timescale in the context of structure functions is subtly different to the definition of Eq.~\ref{eq: tau}.  
We show in Appendix~\ref{sec:app_timescales} that the structure function reports a timescale roughly half that of the method of \citep{OSULLIVAN2024}.\\

We start by modelling both days of the de-trended \textcolor{black}{DRS} RVs with a one-component GP. The results are summarized in Table \ref{tab: GP results}, and the corner plots are shown in Fig. \ref{fig:both night corner plot}. We see that we are able to recover a granulation time scale ($\tau$) to within a 1$\sigma$ of what was predicted using granulation scaling laws \citep{BASU2017}, with the scaling constant taken from Table 2 of \citet{OSullivan2025}, while the variance is within $3\sigma$. The smaller peak in the $\log_{10} \tau $ posterior distribution likely comes from the window function of the data.

\begin{table}
 	\centering
 	\caption{Results for $ \sigma_G$ and $\tau_G $ from different RV data sets corrected for illumination variations. The predicted values are derived from the asteroseismic granulation scaling laws defined in \citet{BASU2017}, with scaling factors derived by \citet{OSullivan2025}.}
 	\label{tab: GP results}
 	\begin{tabular}{ c c  c } 
 		\hline 
 		RVs & $\sigma_G$ [m/s]& $\tau_G $ [minutes] \\
 		\hline
 		Predicted & 0.39 & 35.81  \\
         \hline
        HARPS-N \textcolor{black}{DRS}, 2 nights & $0.23^{+0.05}_{-0.07}$ &  $43.53^{+16.92}_{-13.59}$\\
        \hline 
        HARPS-N \textcolor{black}{DRS}, Night 1  &  $0.28\pm 0.06$ & $44.61^{+19.24}_{-13.99}$ \\
        \hline
        HARPS-N \textcolor{black}{DRS}, Night 2  &  $0.11^{+0.12}_{-0.07}$ & $2111.22^{+3714.78}_{-1667.31}$ \\
        \hline
        HARPS-N \textcolor{black}{DRS}+EXPRES &  $0.21\pm0.05$  &   $44.81^{+15.36}_{-13.59}$\\
        \hline  	
 	\end{tabular}
 \end{table}

To test whether we could obtain the granulation signal with a smaller amount of data (and thus telescope time) we also performed the fit on each night of data individually. These results are shown again in Table \ref{tab: GP results}, with the associated corner plots shown in Appendix \ref{sec:GP Corner Plots}. We see that the signal is detectable on the first night, but not on the second. This is most likely due to the instrumental effects that particularly affected the RVs obtained on the second night, discussed in Section \ref{sec:instrumental} and Appendix \ref{sec:LBL}.  \textcolor{black}{It is worth noting that the timescale is not bimodal in nature when modelling the first night of data independently compared to modelling both nights together. This again is likely due to the instrument systematics, and the reduced effect of the window function when using one night of data compared to two}. Overall, the successful detection of the granulation signal on the first night demonstrates that future studies will require less data, given the optimal observing strategies and conditions.

\textcolor{black}{To rule out any potential impact of p-mode oscillations on the derived granulation properties, we conducted supplementary simulations. The details of this analysis are presented in Appendix \ref{sec:app_timescales} and show that residuals of oscillation signals do not affect the recovery of the granulation signal.}

We also modelled the combined HARPS-N and EXPRES RVs. The results are again shown in Table \ref{tab: GP results}, with the corner plots shown in Appendix \ref{sec:GP Corner Plots}.  A white noise component was incorporated for each instrument in the modelling of the combined dataset. Combining the two datasets neither increased nor decreased the accuracy of the derived variance and timescale, likely because the $RV_{\rm rms}$ for EXPRES is approximately twice that of HARPS-N due to poor seeing on the observing night.

We further used our GPs to analyse the $T_{1/2}$ RVs found by both the \textcolor{black}{$ {\rm CCF}_{T}$} and \textcolor{black}{$ {\rm LBL}_{T}$} methods to see if the signal was more prominent in a certain temperature bin, compared to another. A comparison of the $\log_{10} \sigma_G$ and $\log_{10} \tau_G $ parameters derived is shown in Fig. \ref{fig:CCF_LBL GP results}. \textcolor{black}{From this, we note that in the cooler temperature bins the timescale seems to be more precisely found, especially in the case of the temperature-partitioned CCF RVs. In the hotter $ {\rm CCF}_{T}$ bin, the white noise is notably larger, showing that the GP is struggling to distinguish between the granulation signal and white noise. In the case of the \textcolor{black}{$ {\rm LBL}_{T}$} method, the granulation timescale was also found in the hotter temperature bins, but appears weaker, and the window function peak also appears.}


\begin{figure}
    \centering
    \includegraphics[width=\linewidth]{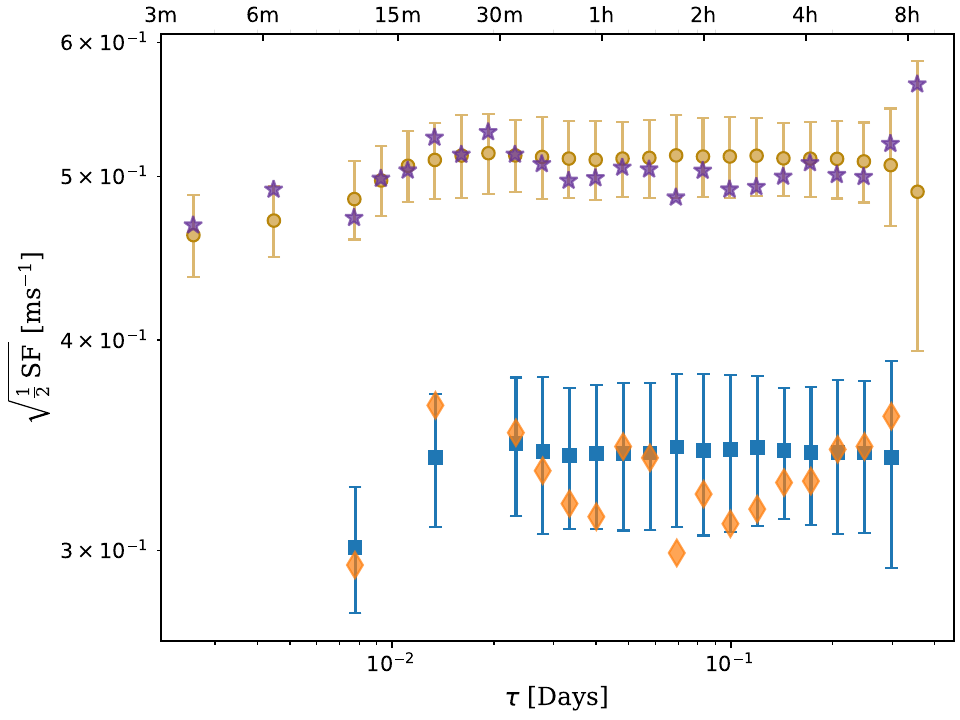}
    \caption{The average of the SFs of the 100 simulated RV time series with cadences of 3.4\,minutes with 0.455~\ms white noise added (yellow circles) and the average of the SFs of the same time series binned in groups of three (blue squares). Error bars represent one standard deviation of the 100 SFs. Structure functions of the HD\,166620 RV data (purple stars) and HD\,166620 RV data binned in threes (orange diamonds) are given for comparison to the simulations. Our model of granulation and white noise has similar features to the HD\,166620 data.}
    \label{fig:SF_GP+WN}
\end{figure}

\subsubsection{Structure function analysis of GP simulations}
\label{sec:SF_GP_simulations}

Both the structure function and GP analyses detect correlated noise with an amplitude of $\sim20$\cms at $\simeq20$ minutes on the SF definition of timescale, or $\simeq40$ minutes using the GP definition (we show that these two timescales are equivalent in Appendix~\ref{sec:app_timescales}).
To test whether these signals really are equivalent
we calculated SFs of RVs simulated using the GP hyper parameters with added white noise. The simulated RVs were calculated using  {\tt celerite2}'s `sample' package to simulate a time series with a given set of hyper-parameters. We used the GP hyper parameters found by using both nights of the detrended RVs.

We take 100 realisations of GP RV simulations at a cadence of 1~s which are then binned to the same cadence as the HD\,166620 data (3.4\,minutes). We then add white noise with a standard deviation of 0.454\,\ms to each simulation. As can be seen in Fig.~\ref{fig:SF_GP+WN}, we are able to reproduce a similar relative fall in $\sqrt{\nicefrac{1}{2}\,{\rm SF}}$ as we observe for HD\,166620 when we bin the RV data in groups of three. The SF generally has a small slope leading into timescales of around 23 minutes. We also tested adding additional correlated noise to the model (to model a potential instrumental signal), however, the calculated SFs have an extremely similar profile except for a slightly smaller drop in the binned-in-threes SF. From this, we conclude that the RVs for HD\,166620 are likely made from uncorrelated noise and correlated noise such as granulation. We are not able to conclude whether the uncorrelated noise is instrumental or astrophysical.

\begin{figure}
    \centering
    \includegraphics[width=\linewidth]{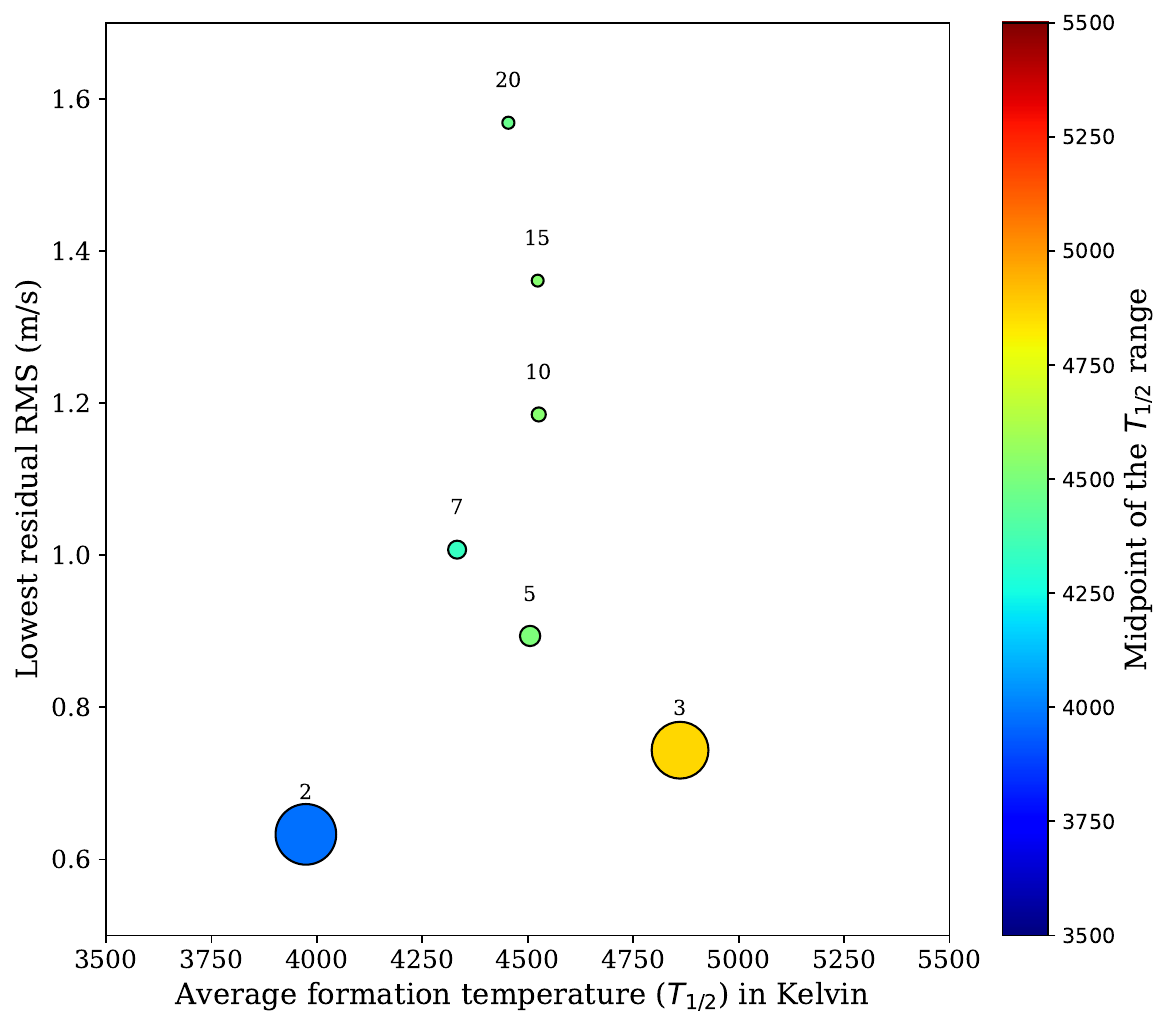}
    \caption{ \textcolor{black}{Lowest RMS residuals from ${\rm CCF}_{T}$ RVs after instrumental correction, obtained across different temperature binning strategies. Numeric labels above the points denote the number of bins used in each case. Each point represents the minimum RMS achieved for a given number of bins, with point size reflecting the width of the corresponding temperature bin and colour indicating its midpoint. While the minimum RMS tends to increase with the number of bins (i.e., finer binning), it is consistently found in medium-temperature bins, regardless of bin size or total bin count.}}
    
    \label{fig:RMS_CCF_LBL}
\end{figure}

    




\subsection{Temperature dependence of RV RMS}

We find that the residual root-mean-squared (RMS) velocities measured by the temperature-partitioned \textcolor{black}{$ {\rm CCF}_{T}$} method show a distinctive high-low-high dependence on the $T_{1/2}$. \textcolor{black}{In other words, we observed higher RMS values in the extreme temperature bins when compared to the intermediate temperature ranges.  Fig. \ref{fig:app_RMS} presents the high-low-high trend in the RMS values calculated from 20 temperature bins, each optimized for balanced RV variance.} The high-low-high pattern does not appear to have originated from the CCD illumination correction, as the raw RVs (semi-transparent circles) also exhibit the same pattern. \textcolor{black}{The circle sizes are scaled according to bin sizes.}

\textcolor{black}{To assess whether this feature depends on bin size and/or bin count, we generated $n$ variance-balanced bins (with $n$ ranging from 2 to 20) and extracted RVs using the \textcolor{black}{$ {\rm CCF}_{T}$} method.
Fig. \ref{fig:RMS_CCF_LBL} illustrates the lowest RMS recorded among different $T_{1/2}$ ranges for varying bin counts from 2 to 20.  In all cases, the residual $RV_{\rm rms}$ is relatively larger at the coldest and hottest $T_{1/2}$ extremes, while the lowest RMS consistently occurs at intermediate temperatures within the line-forming layers of the photosphere. This pattern persists irrespective of bin size or bin count.}

\textcolor{black}{One might wonder whether the low RMS simply reflects averaging over more spectral lines. Yet, because since the bins are balanced for RV content and scatter, the observed high-low-high trend is unlikely to stem solely from a signal-to-noise effect. That said, RMS remains sensitive to other sources of variability—such as instrumental noise, stellar activity, and systematic effects—which may influence the overall amplitude while preserving the underlying trend.}

Before drawing any conclusions, we must verify whether the high-low-high RV dispersion pattern is astrophysical (i.e., a function of $T_{1/2}$) and not a function of bin strength. To do this, we re-calibrated the bins to have comparable mask weights and balanced the weighted variance across the bins. We experimented with different numbers of bins (from 3 to 20) and found no strong dependency between the $RV_{\rm rms}$ and the cumulative mask weights (line strengths) in each bin. Interestingly, the high-low-high RMS pattern persisted in the mask weight-balanced bins, further suggesting that these variations are most likely astrophysical and not dominated by noise scaling. 

\textcolor{black}{A similar high-low-high pattern is apparent in simulations made by \citet{Beeck2013} using the 3D radiative magnetohydrodynamic code MURaM \citep{Vogler2003,Vogler2005}. The lower right-hand panels of their Fig. 3 show the flow vectors and temperature stratification across a granule boundary in a K5V model, which is somewhat cooler than HD 166620. The scatter in vertical velocity clearly decreases with decreasing temperature, passing through a minimum at temperatures between 3000K and 4000K, and increasing slightly in higher, cooler layers. In the Sun, strong lines whose cores form in these highest layers also show RV variations that are anticorrelated with those emanating from deeper, hotter layers \citep{AlMoulla2022}. The RMS scatter of individual line RVs must therefore pass through a minimum between these two extremes.}

\section{Discussion}
\subsection{Sensitivity across photospheric depths}
By examining the spectra and RVs as a function of $T_{1/2}$ bins, we investigated the sensitivity of granulation-induced RV variations across different photospheric layers. 
\textcolor{black}{We found that granulation timescales were more precisely detected in the RVs derived from the cooler bins using the {$ {\rm CCF}_{T}$} method. In contrast, the hotter {$ {\rm CCF}_{T}$} bins appeared dominated by the window function, even though they contain comparable RV information. This could suggest that the shallower layers in the photosphere of this magnetically quiet K dwarf may be better suited for detecting granulation RV signals than the deeper layers.}

The third signature of granulation \citep{Gray2009}, the net observed convective blueshift, is known to depend on line depths. Weak lines are expected to exhibit a stronger blueshift \citep[][]{Ramirez2008, Dravins1981, Dravins1986} as they form deeper in the photosphere, where the correlation between intensity and velocity fields due to granulation is most pronounced. In their study of the Sun, \citet{AlMoulla2022} and \citet{Ellwarth2023} also observed that convective blueshift decreases with reduced temperatures, suggesting that the blueshift could be less pronounced in the upper layers. \textcolor{black}{Our analysis of the K-dwarf HD~166620 suggests that granulation-induced RV variations in the cooler temperature bin, corresponding to the upper layers, align more closely with predicted values. However, the signal does seem to be present in the hotter bins as well, although it is less precisely characterised and more of the signal is attributed to white noise. }
\\
We propose that this divergence from solar-based investigations may arise from the following factors:

1) \textcolor{black}{While the two nights are not well-sufficient to detect supergranulation, Fig. \ref{fig:CCF_LBL GP results} shows that we see timescales similar to the expected supergranulation signals in addition to the granulation signal in the hotter bin (deeper photospheric layers). This could also be attributable to the window function, but since we are employing a one-component GP model, either of these factors could potentially influence the detectability (convergence) of the granulation signal.}

2) Granulation properties strongly depend on stellar fundamental properties like the $\rm T_{eff}, log \textit{g}$ and [Fe/H]. The literature \citep[e.g. ][]{Samadi2013b, Samadi2013a} indicates that there have not been enough studies on K-dwarfs to update the existing granulation models and scaling relations accurately. This is the first time we have analysed the photospheric depth dependence of the observable granulation signature on a K-dwarf.




\section{Conclusion}

This paper presents a pilot study aimed at detecting and characterising the granulation signature on a star other than the Sun, using HARPS-N. We conducted dense observations of the quiet K-dwarf, HD 166620, over two consecutive nights to capture granulation variability with timescales ranging from minutes to hours. On the first night, we also obtained contemporaneous observations from EXPRES.

Our analysis revealed significant instrument-induced radial velocity (RV) variations in the HARPS-N data, on the order of a few \ms. These variations were traced back to exceptionally good seeing conditions during the HARPS-N observations. \textcolor{black}{Under such stable conditions, the existing scrambling proved insufficient, leading to fibre injection instabilities.} Consequently, this led to variations in CCD illumination, resulting in spurious RV measurements (upto 2 \ms) across the CCD. \textcolor{black}{Given that many EPRV spectrographs share similar fibre specifications, we emphasize that such instrumental effects under ideal conditions are likely not unique to HARPS-N.} Due to suboptimal seeing conditions on the nights of observation at the EXPRES site, the EXPRES observations were less affected by these spurious RV variations.

Analysing the instrument-corrected HARPS-N RVs using a one-component GP granulation model, we found the granulation signal to have a characteristic timescale ($\tau_{G}$) of  $43.53^{+16.92}_{-13.59}$ minutes and a standard deviation ($\sigma_{G}$) of $23^{+5}_{-7}$ cm/s. However, we found no evidence of supergranulation, which could be attributed either to the inadequate baseline of observations or to the fundamentally different nature of the RV time series.

The structure function analysis also yields concordant results, suggesting our conclusions are not strongly dependent on the analysis techniques.

To assess the relative susceptibility of spectral line segments formed at different photospheric depths to granulation, we analysed RVs as a function of the average line formation temperature ($T_{1/2}$). We generated two distinct RV time series corresponding to unique formation temperature bins with comparable RV content and variance, ranging from the coolest to the hottest. For context, RVs extracted from spectral regions formed at cooler temperatures correspond to the upper layers of the photosphere, and vice versa. 
To mitigate potential biases in RV extraction, we employed both the \textcolor{black}{$ {\rm CCF}_{T}$} and \textcolor{black}{$ {\rm LBL}_{T}$} methods in parallel. 
Using GP analysis, we examined the two $T_{1/2}$-sensitive RV time series ($\times$2) for granulation signatures.
\textcolor{black}{We found that detections in the cooler temperature bins more reliably match the values predicted by the scaling laws, while those in the hotter bins tend to be dominated by the window function.} 
\textcolor{black}{Although non-astrophysical factors could contribute to this effect, our findings suggests that granulation-induced variations are more detectable in stronger spectral lines formed in the shallower layers of the photosphere.} This holds even when accounting for the difference in the information content within each $T_{1/2}$ bin. 
Our \textcolor{black}{$ {\rm CCF}_{T}$} and \textcolor{black}{$ {\rm LBL}_{T}$} results agree within one sigma of each other.

This study quantitatively characterises the RV signature of granulation on a quiet K-dwarf, representing a significant step towards improving RV precision for detecting low-mass exoplanets.
Future work will broaden the scope by extending these studies to stars of other spectral types as well as to supergranulation studies. Furthermore, this research highlights the pivotal role of occasional, targeted, and dense observation campaigns in uncovering critical insights into stellar behaviour and instrument performance.





\section*{Acknowledgements}
\textcolor{black}{AAJ is grateful to the MNRAS reviewer for their constructive feedback, which significantly improved the scientific quality of this work.}
AAJ, AM, and BSL acknowledge funding from a UKRI Future Leader Fellowship, grant number MR/X033244/1. AM acknowledges funding from a UK Science and Technology Facilities Council (STFC) small grant ST/Y002334/1. KA acknowledges support from the Swiss National Science Foundation (SNSF) under the Postdoc Mobility grant P500PT\_230225.  This work has been carried out within the framework of the NCCR PlanetS supported by the Swiss National Science Foundation under grants 51NF40\_182901 and 51NF40\_205606. NKOS, MC, and SA acknowledge funding from the European Research Council under the European Union’s Horizon 2020 research and innovation programme (grant agreement no. 865624, GPRV). C.A.W. would like to acknowledge support from the UK Science and Technology Facilities Council (STFC, grant number ST/X00094X/1).
X.D acknowledges the support from the European Research Council (ERC) under the European Union’s Horizon 2020 research and innovation programme (grant agreement SCORE No 851555) and from the Swiss National Science Foundation under the grant SPECTRE (No 200021\_215200). 
JF acknowledges support from STFC grant number ST/Y509383/1. HMC acknowledges funding from a UKRI Future Leader Fellowship (grant numbers MR/S035214/1 and MR/Y011759/1). 

\textcolor{black}{This work is based on observations made with the Italian Telescopio Nazionale Galileo (TNG) operated on the island of La Palma by the Fundaci\'on Galileo Galilei of the INAF (Istituto Nazionale di Astrofisica) at the Spanish Observatorio del Roque de los Muchachos of the Instituto de Astrofisica de Canarias. The HARPS-N project was funded by the Prodex Program of the Swiss Space Office (SSO), the Harvard University Origin of Life Initiative (HUOLI), the Scottish Universities Physics Alliance (SUPA), the University of Geneva, the Smithsonian Astrophysical Observatory (SAO), the Italian National Astrophysical Institute (INAF), University of St. Andrews, Queen’s University Belfast, and University of Edinburgh.}

The data presented here made use of the Lowell Discovery Telescope at Lowell Observatory. Lowell is a private, non-profit institution dedicated to astrophysical research and public appreciation of astronomy and operates the LDT in partnership with Boston University, the University of Maryland, the University of Toledo, Northern Arizona University and Yale University. We gratefully acknowledge ongoing support for telescope time from Yale University, the Heising-Simons Foundation, and an anonymous donor in the Yale community. We especially thank the NSF for funding that allowed for precise wavelength calibration and software pipelines through NSF ATI-1509436 and NSF AST-1616086 and for the construction of EXPRES through MRI-1429365. \textcolor{black}{This work has made use of the VALD database, operated at Uppsala University, the Institute of Astronomy RAS in Moscow, and the University of Vienna.}
\addcontentsline{toc}{section}{Acknowledgements}


\section*{Data Availability}
The RV data will be made available via Vizier CDS. \textcolor{black}{This work makes use of the HARPS-N solar RVs, which are available at \url{https://doi.org/10.82180/dace-h4s8lp7c} and will be described in Dumusque et al. 2025 in prep.  The SDO/HMI images are publicly available at \url{https://sdo.gsfc.nasa.gov/data/}}
 




\bibliographystyle{mnras}
\bibliography{Bible}



\appendix

\section{P-Mode Simulations}
\label{sec:p-mode}
\textcolor{black}{ The p-mode oscillation timescale in the K-dwarf HD\,166620 is expected to be shorter than that of the Sun \citep{Chaplin2019}. We adopted 180-second exposures to average out these variations. To verify that this duration sufficiently suppresses the p-mode signal, we have run simulations to make sure any signal in our data related to p-mode does not affect the granulation properties we derive. To do this, we simulate RVs containing the granulation signal, signal from the p-modes, and white noise. The p-modes were simulated using a periodic SHO term with Q = 5.66, S = $5.54 \times 10^{-5}$ $m^2 s^{-2} days$, and w = 0.027 $m/s$, corresponding to a timescale of 3.98  minutes. We derive these values using scaling laws \citep{BASU2017} and the values for the p-modes given in \citet{AlMoulla2023}. The granulation and white noise are simulated using the same method as discussed in Section \ref{sec:SF_GP_simulations}. We then modelled the data using the same method as we used for the actual RVs. To account for the effect of the exposure times, we first simulated a time series with an observation every second during the HARPS-N observing night and binned this down to 180 seconds. We ran this simulation 5 times and the results are shown in Fig .\ref{fig:p_mode sim}. These show that the granulation standard deviation and timescale are correctly recovered, even when not explicitly modelling for the p-modes. }

\begin{figure}
    \centering
    \includegraphics[width=0.48\textwidth]{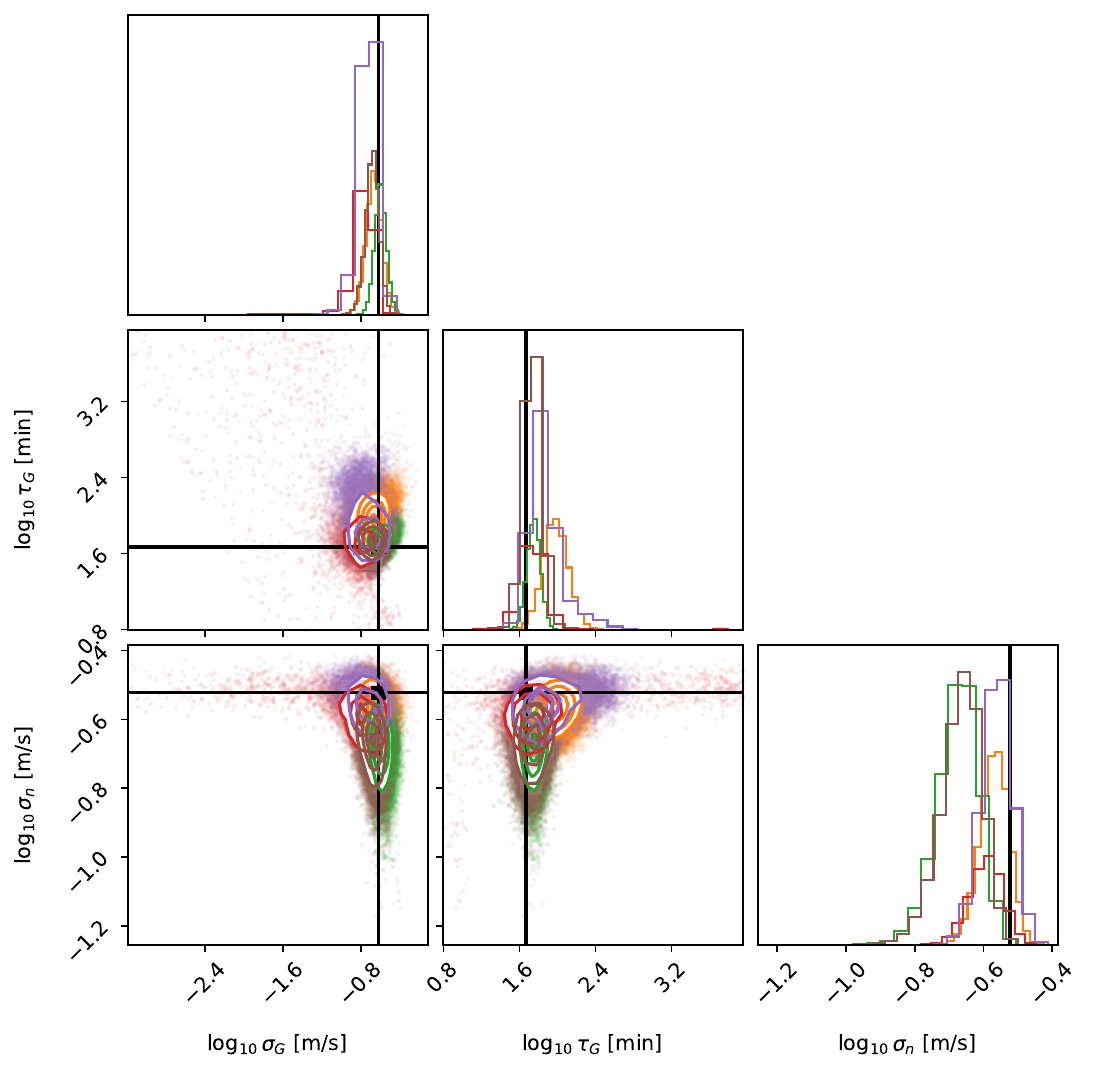}
    \caption{\textcolor{black}{MCMC posterior distribution plots for the one-aperiodic GP model of simulated RVs, including p-mode, granulation, and white noise. The parameters shown are the granulation standard deviation, $ \log_{10}\sigma_G$ , the granulation timescale, $\log_{10}\tau_G$ , and the white noise, $\log_{10}\sigma_n$. The 1-D posterior distributions for each parameter, marginalised over all the other parameters, are shown by the histograms in the diagonal panels. The predicted values are indicated by the black lines. The smaller peak in the distributions likely comes from the window function. The 2-D posteriors are shown in the off-diagonal panels. }}
    \label{fig:p_mode sim}
\end{figure}

\section{Timescales from Gaussian Processes and from structure functions} \label{sec:app_timescales}

In Section \ref{sec:results} we introduce two definitions of timescale.
The first is defined as the knee in the structure function, beyond which observations separated by longer lags show no increase in variability.
The second is defined as $\tau = \frac{1}{\nu_0}$ where $\nu_0$  satisfies
\begin{equation}
    P_\mathrm{H} (\nu = \nu_0) = \frac{1}{2}P_\mathrm{H}(\nu = 0),
\end{equation}
\citep[see][]{OSULLIVAN2024}.
To compare timescales from these two analysis methods, we simulate a time series with a granulation and supergranulation component following the method of \citet{OSULLIVAN2024}.
In Fig .\ref{fig:sf_timescale_test} we compare the two definitions of timescale and find that the structure function timescale is roughly half of the timescale defined in the Gaussian Process.
\begin{figure}
    \centering
    \includegraphics[width=1\linewidth]{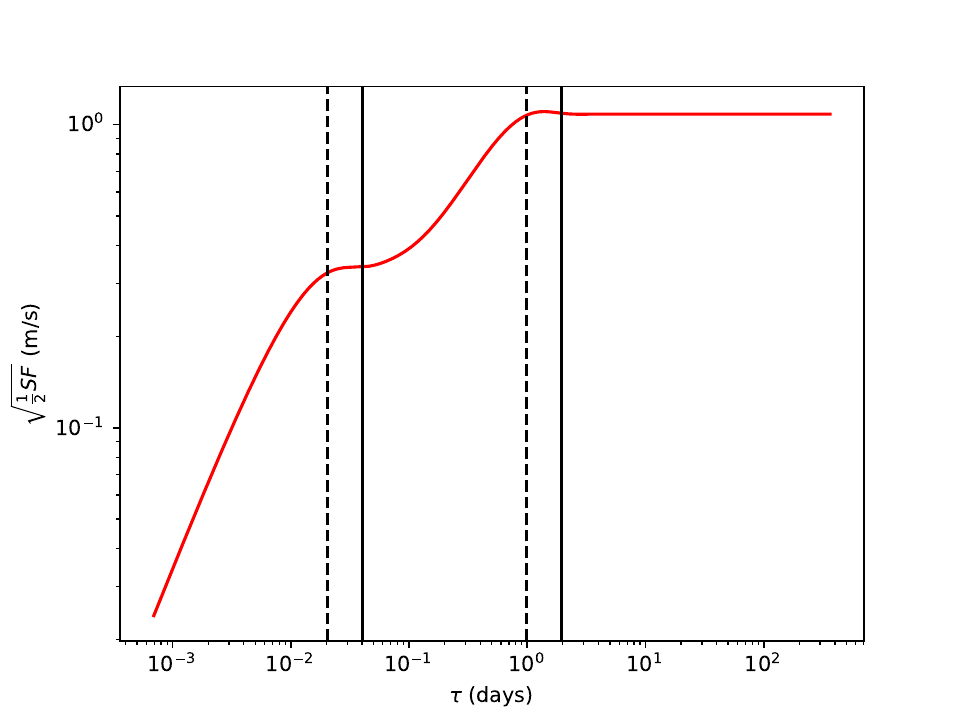}
    \caption{Structure function of a time series generated from a two-component Gaussian Process. Black solid lines indicate the timescales of each component as defined in \citet{OSULLIVAN2024} and dashed lines show half their value. }
    \label{fig:sf_timescale_test}
\end{figure}

\section{Diagnostics of instrumental systematics}
\label{sec:step}

We observed additional features in the lower panel of Fig. \ref{fig:DiagnosticRatios}, which were not explained by airmass alone.
These features seem to dominate the radial velocities extracted from the cooler regions of the spectra (Fig. \ref{Day1steps}), as detailed in Section \ref{sec:temp_formalism}.
\begin{figure}
    \centering
    \includegraphics[width=0.75\linewidth, angle=270]{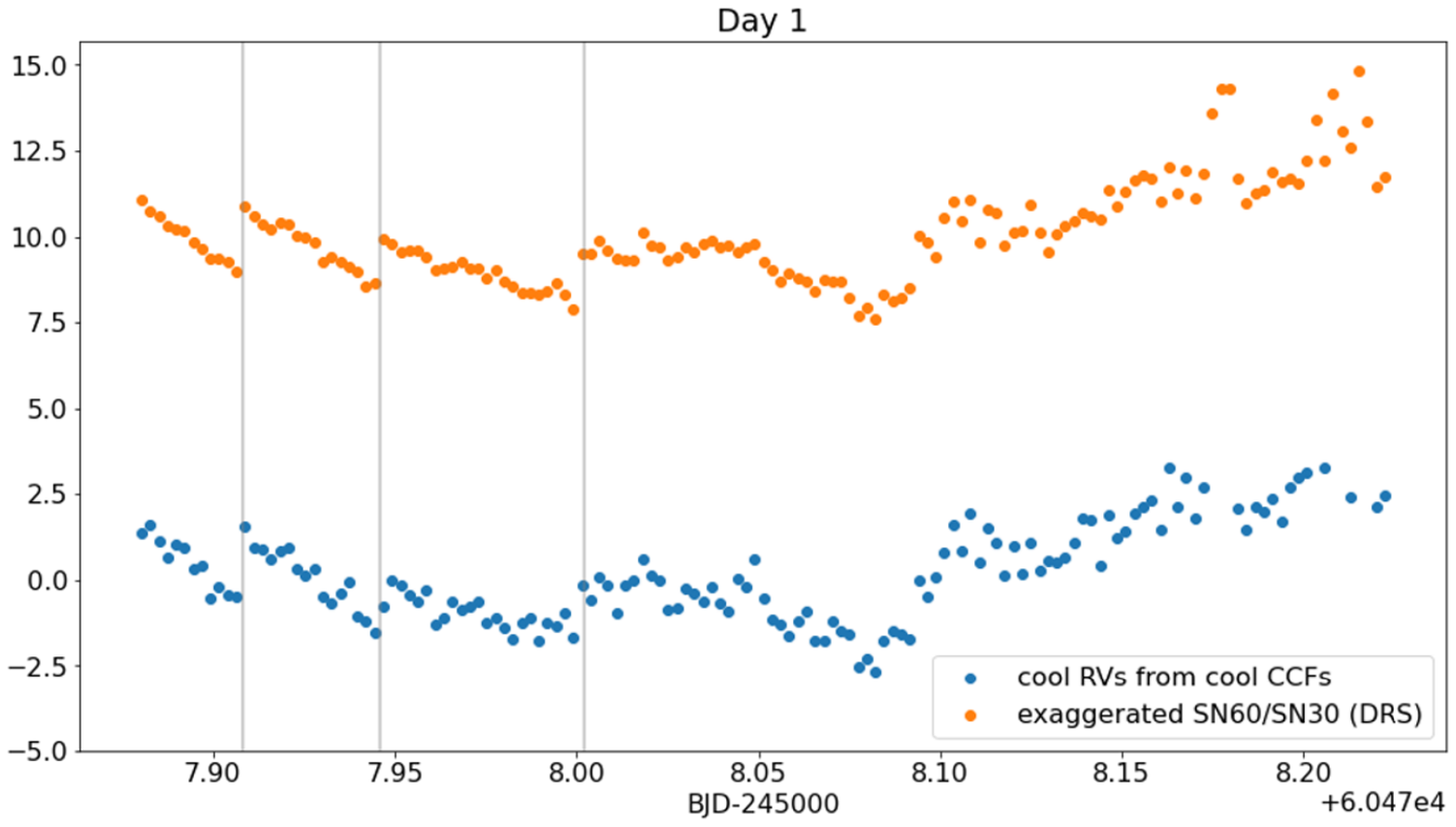}
    \caption{ The RV time series extracted from the cooler part of the spectra for day 1 is shown, illustrating the same steps in the SNR and the corresponding residuals in the bottom panel of Fig. \ref{fig:DiagnosticRatios}. A similar match was observed for day 2 also.}
    \label{Day1steps}
\end{figure}

\section{Line-by-line RVs over the detector}
\label{sec:LBL}

To determine whether the hour timescale signal observed during the two nights is caused by the star or by instrumental factors, we referred to the findings of \citet{Bouchy2013}. Their observation suggested that issues with fibre injection can create a significant signature spanning from left to right on the detector. Previous studies have already shown that LBL RVs are a powerful tool for instrumental diagnostic systematics (such as the ageing of the ThAr lamp as demonstrated in \citet{Cretignier2023} or the stitching of the HARPS CCD in \citet{Cretignier2021}) due to their ability to extract local RV information.

We extracted the LBL RVs following \citet{Dumusque2018} on the S2D spectra corrected by YARARA \citep{Cretignier2021}. By doing so, some extra signals such as the ones induced by tellurics lines or instrumental point-spread-function variation \citep{Stalport2023} can already be mitigated from the LBL RVs. The line selection used for the LBL RVs is the line selection tailored to HD166620 and obtained from the master spectrum deliver by YARARA. The optimal line selection was obtained as described in \citet{Cretignier2020} in order to achieve the optimal extraction and reach the highest SNR.

We show in Fig.\ref{FigLBLDetector} and Fig.\ref{FigLBLDetector2} the LBL RVs binned in a 2 $\times$ 4 mosaic over the detector (1 split in cross-dispersion and 3 along the main order) for each night individually. The echelle-order spectrum with the same subdivisions is shown in Fig.\ref{FigLBLDetector0}. The blue part of the detector is displayed in the top row, while the red part is stacked in the bottom row. Although no clear chromatic effect is visible by comparing the two rows, a large left-to-right effect is detected, which exhibits a signature very similar to the one observed on the DRS RVs (top panel of Fig. \ref{fig:DiagnosticRatios}). 

In practice, such an instrumental signature could be very well corrected by the LBL PCA presented in \citet{Cretignier2023}. However, it is not clear if the PCA would also absorb the stellar activity signals we are aiming to detect and characterize. This is notably true, in particular, if granulation is expected to affect some lines more than others (as is the case; \citealt{Reiners2016, Meunier2017}). For this reason, we did not opt for such a correction in the paper and rather relied on the CCD lightcurves (see Sect.~\ref{sec:instrumental}) to provide the time-domain correction, since less granulation signatures is expected from the CCD photometry in comparison to LBL RVs. 

Since this new instrumental signature is being identified for the first time on HARPS-N in this work, we also assess its expected contribution on the full lifetime of the instrument. We carried out the same LBL PCA analysis on historical observations of the star. For this, we used data from two years preceding the current campaign, obtained as part of the RPS program. Our corrected LBL RVs by the PCA reveal that this fibre injection may result in a 46 \cms jitter along the lifetime of the instrument. Naturally, this result should be carefully considered since it is not clear if the PCA over the two years is correcting for the same instrumental effects as the one obtained over the two nights. This is particularly true given that the seeing was exceptionally good for the two nights and the observing conditions were therefore very different from regular observations.

However, if verified, this element represents non-negligible contamination in the instrumental RV budget that needs to be investigated, in particular when planetary signals smaller than 30 cm/s are chased \citep{Hall2018}. 
\begin{figure}
    \centering
    \includegraphics[width=1\linewidth]{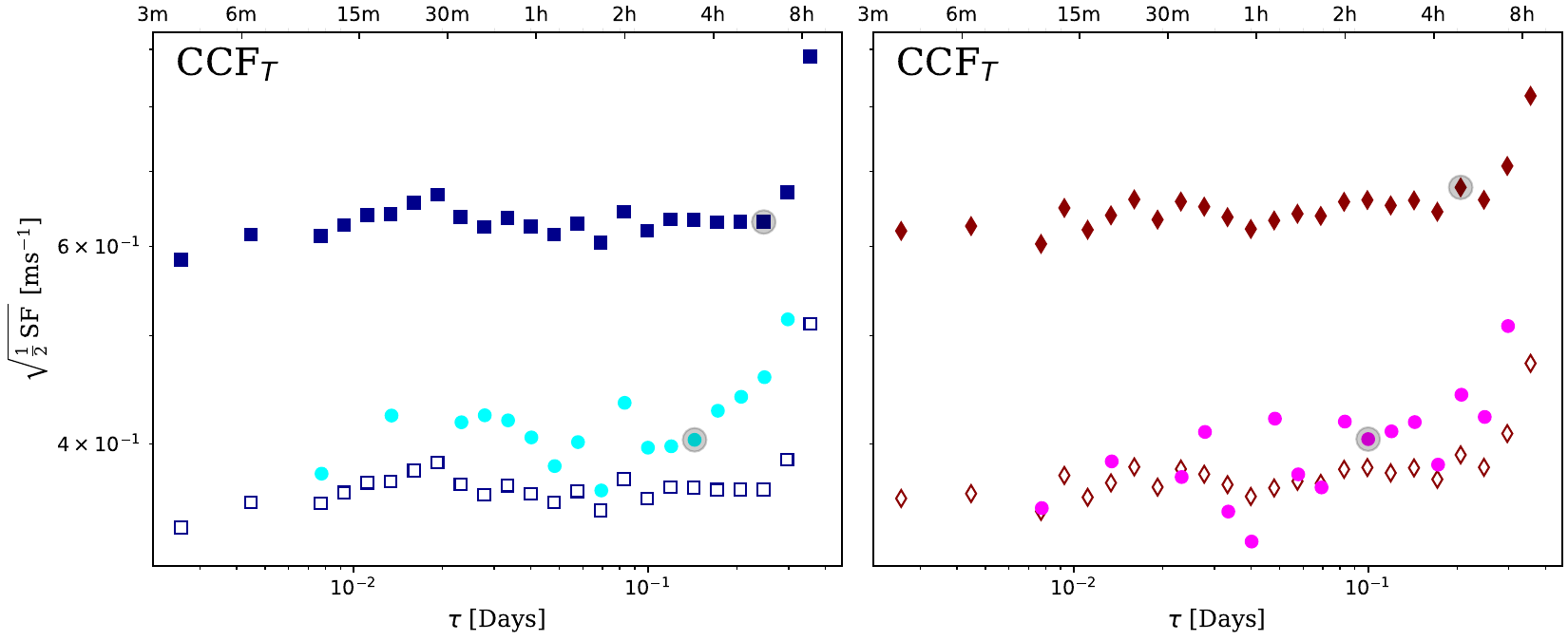}
    \caption{\textcolor{black}{$\sqrt{\nicefrac{1}{2}\,{\rm SF}}$ of the ${\rm CCF}_{T}$ binned radial velocities as a function of timescale. The filled squares and diamonds give the structure functions of the cool and hot bin ${\rm CCF}_{T}$ radial velocities respectively as in Figure~\ref{fig:SF_temp_bins}. The unfilled squares and diamonds give the structure functions divided by $\sqrt{3}$ to show where a time series of purely uncorrelated noise would fall to when binned by groups of three. The cyan and magenta circles give the structure function of the binned-in-threes radial velocities for the cooler and hotter bins respectively to extract the correlated and uncorrelated noise amplitudes. The over plotted grey circles give the last point used to calculate the average RMS. Note that the structure function of the hot bin radial velocities binned in threes overlay the original hot bin structure function divided by $\sqrt{3}$ implying that there is less correlated signal in the ${\rm CCF}_{T}$ hot bins than any of the other temperature dependent datasets.}}   
    \label{fig:SF_CCF_threes}
\end{figure}
\begin{figure*}
    \centering
    \includegraphics[width=\linewidth, angle=0]{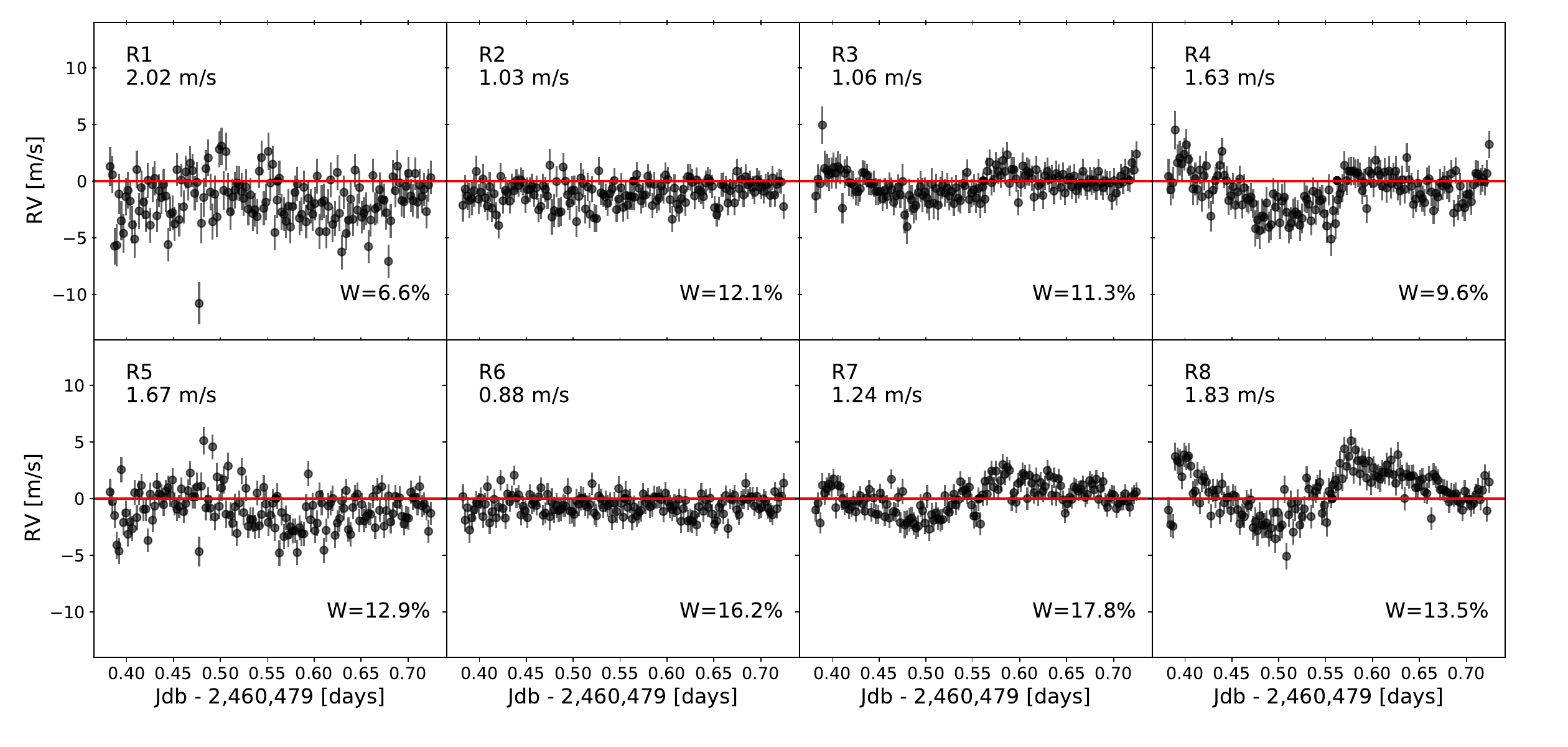}
    \caption{Line-by-line RVs binned regions over the 2 $\times$ 4 mosaic of the HARPS-N detector (cross-dispersion is along the vertical axis, main order along the horizontal one). Only the second night of the campaign with the large modulation is displayed. Top left panel (R1) represents the left blue region of the HARPS-N detector, the bottom right panel (R8) the right red part of the detector. The RV standard deviation of each region is indicated in each panel as well the relative contribution weight of the region to the RV DRS measurement. A clear left-to-right signature (along the main order direction) is visible with a small chromatic effect (along the cross-dispersion). }
    \label{FigLBLDetector}
\end{figure*}

\begin{figure*}
    \centering
    \includegraphics[width=\linewidth, angle=0]{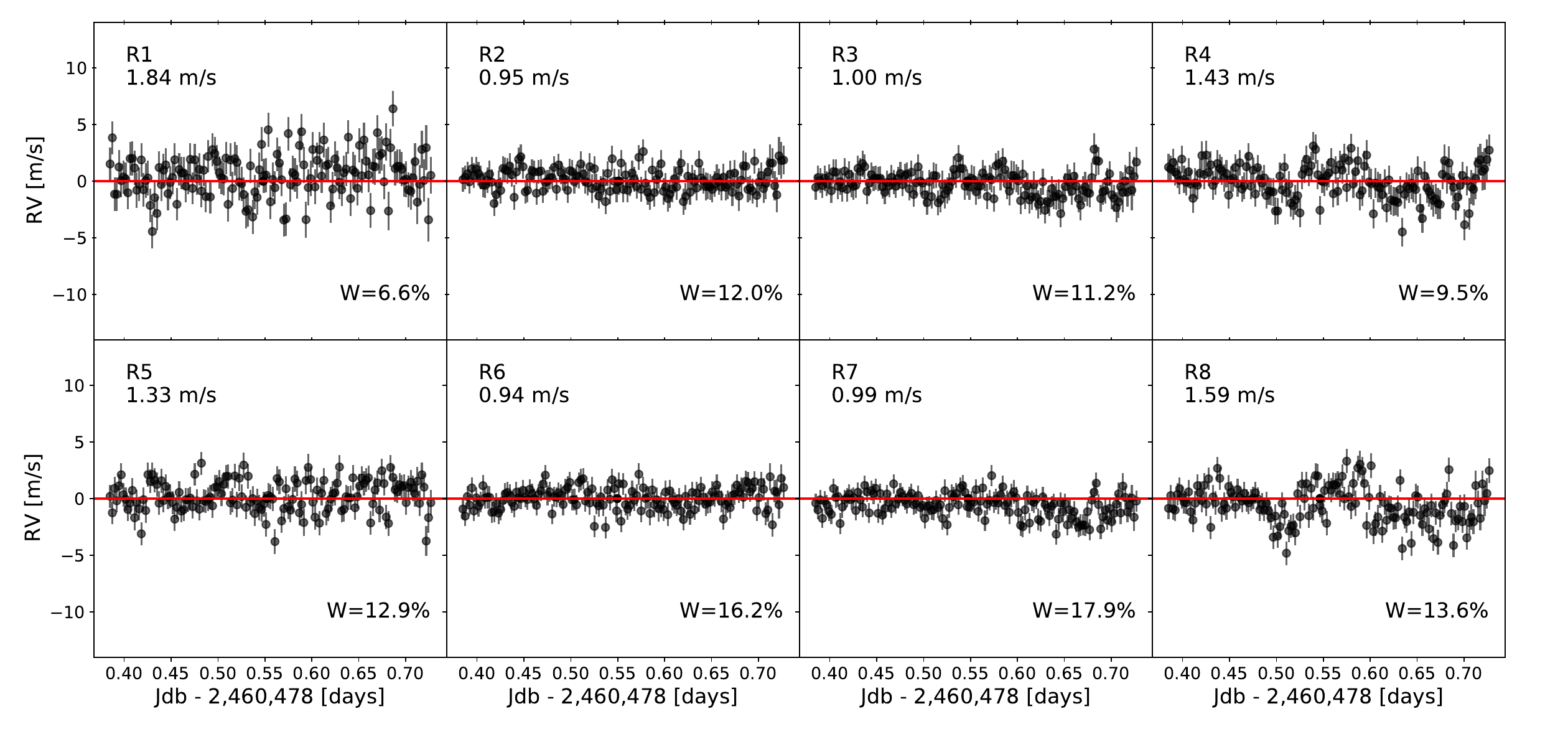}
    \caption{Same as Fig.\ref{FigLBLDetector2} for the first night of the campaign. Again some modulation identified in the DRS RVs are mainly instrumental.}
    \label{FigLBLDetector2}
\end{figure*}

\begin{figure*}
    \centering
    \includegraphics[width=\linewidth, angle=0]{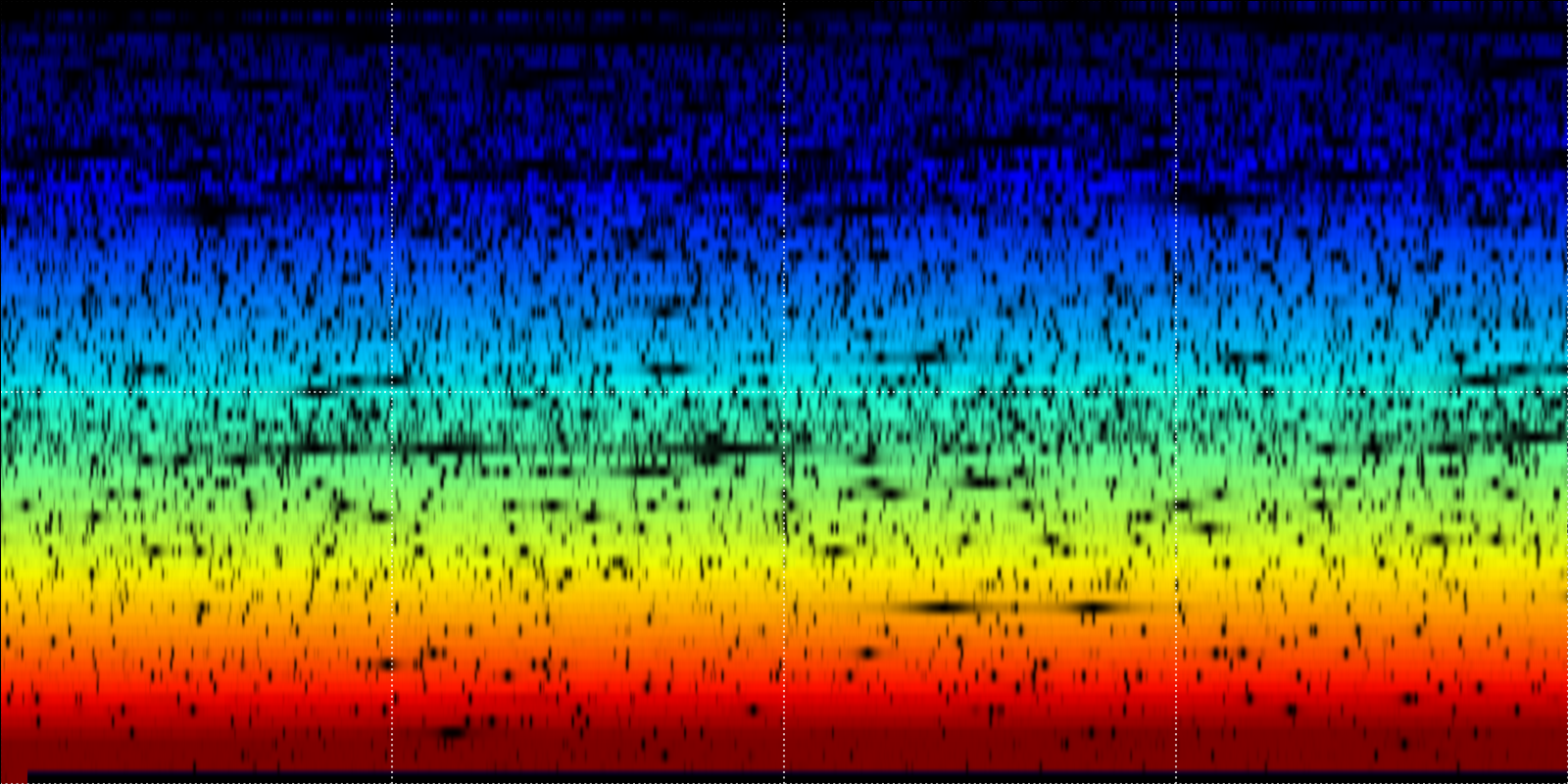}
    \caption{Representation of the HARPN echelle-order spectrum of HD166620 and the 2 $\times$ 4 mosaic used to bin the LBL RVs in Fig.\ref{FigLBLDetector} and Fig.\ref{FigLBLDetector2}. Absorption stellar lines are visible as dark features.}
    \label{FigLBLDetector0}
\end{figure*}

\section{RV time series as a function of $T_{1/2}$}
To investigate the sensitivity of spectral regions formed at different photospheric depths, to variations induced by granulation, we measure the RV as a function of the average formation temperature, $T_{1/2}$. Fig .~\ref{fig:28CCFRVs} shows the 2 (cold \& hot ) $\times$ 2 (\textcolor{black}{$ {\rm CCF}_{T}$} \& \textcolor{black}{$ {\rm LBL}_{T}$}) RV timeseries after correcting for the instrumental systematics discussed in Sect.~\ref{sec:instrumental}.



\begin{figure*}
    \centering
    \includegraphics[width=0.48\linewidth]{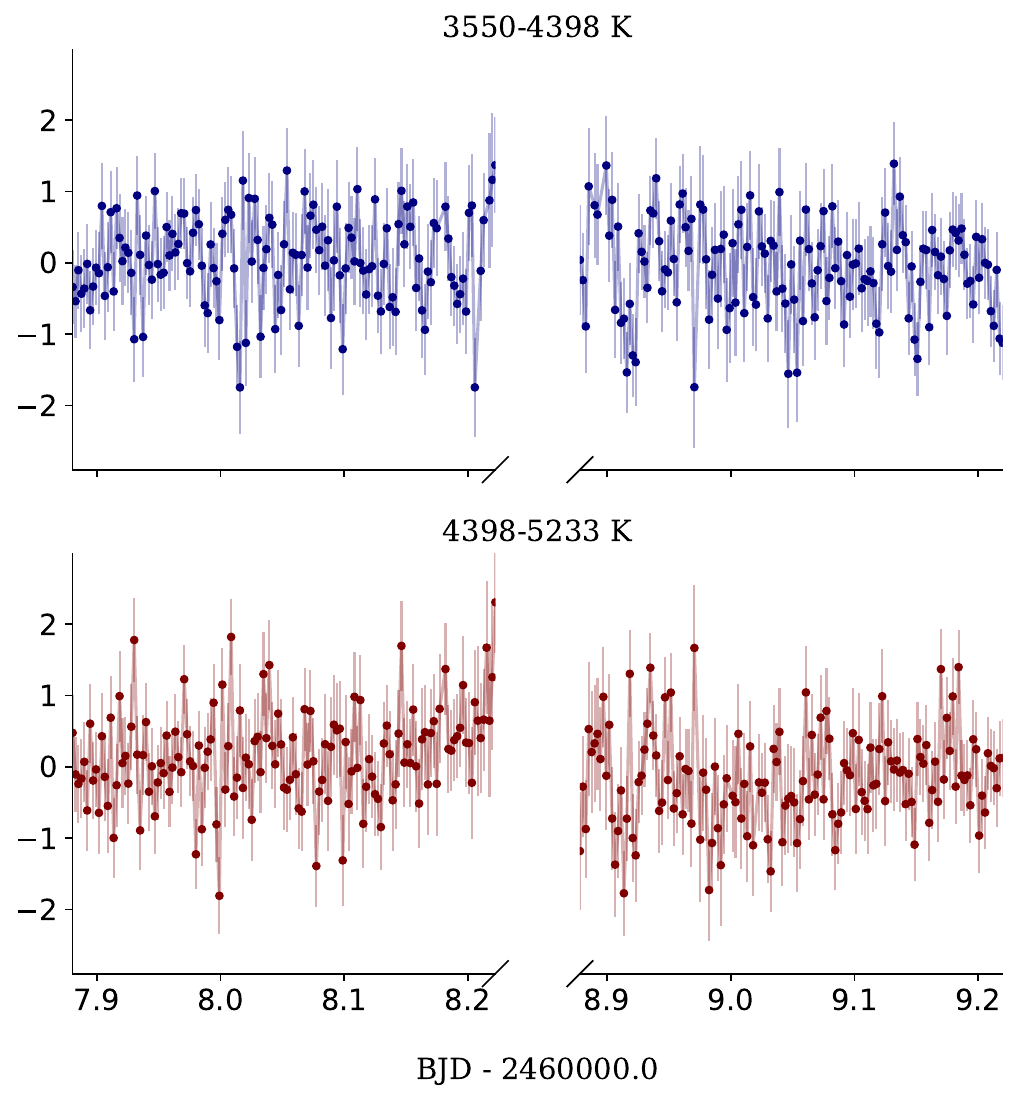}\hfill
    \includegraphics[width=0.48\linewidth]{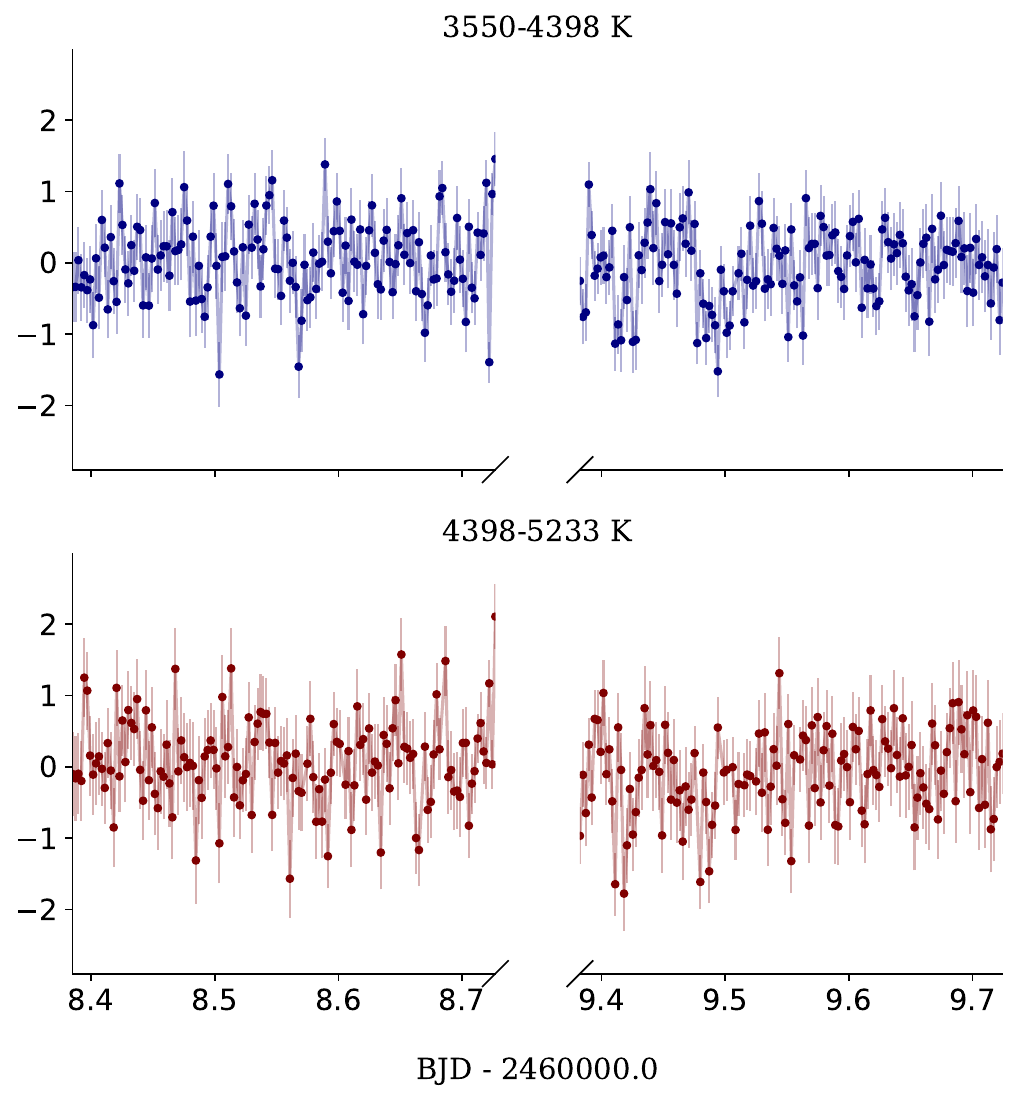}
    \caption{\textcolor{black}{$ {\rm CCF}_{T}$ (left) and {$ {\rm LBL}_{T}$} (right)} RV time series for the 2 temperature ranges described in Sect.~\ref{sec:temp_ranges}. These RVs are corrected for the CCD illumination variations as described in Sect.~\ref{sec:instrumental}.}
    \label{fig:28CCFRVs}
\end{figure*}
\section{GP Corner Plots}
\label{sec:GP Corner Plots}
We performed the GP fit on each night of data individually. These results are shown as corner plots in Figs. \ref{fig:night 1 corner plot} and \ref{fig:night 2 corner plots}, with the associated values in Table \ref{tab: GP results}. We see that the signal is detectable on the first night, but not on the second. This is most likely due to the instrumental effects that particularly affected the RVs obtained on the second night, discussed in Section \ref{sec:instrumental} and Sect. \ref{sec:LBL}.

We also modelled the combined HARPS-N and EXPRES RVs. The results are again shown in Table \ref{tab: GP results}, with the corner plot shown in Fig.  \ref{fig:H+E}. Combining the two datasets neither increased nor decreased the accuracy of the derived variance and timescale. 

\begin{figure}
    \centering
    \includegraphics[width=0.48\textwidth]{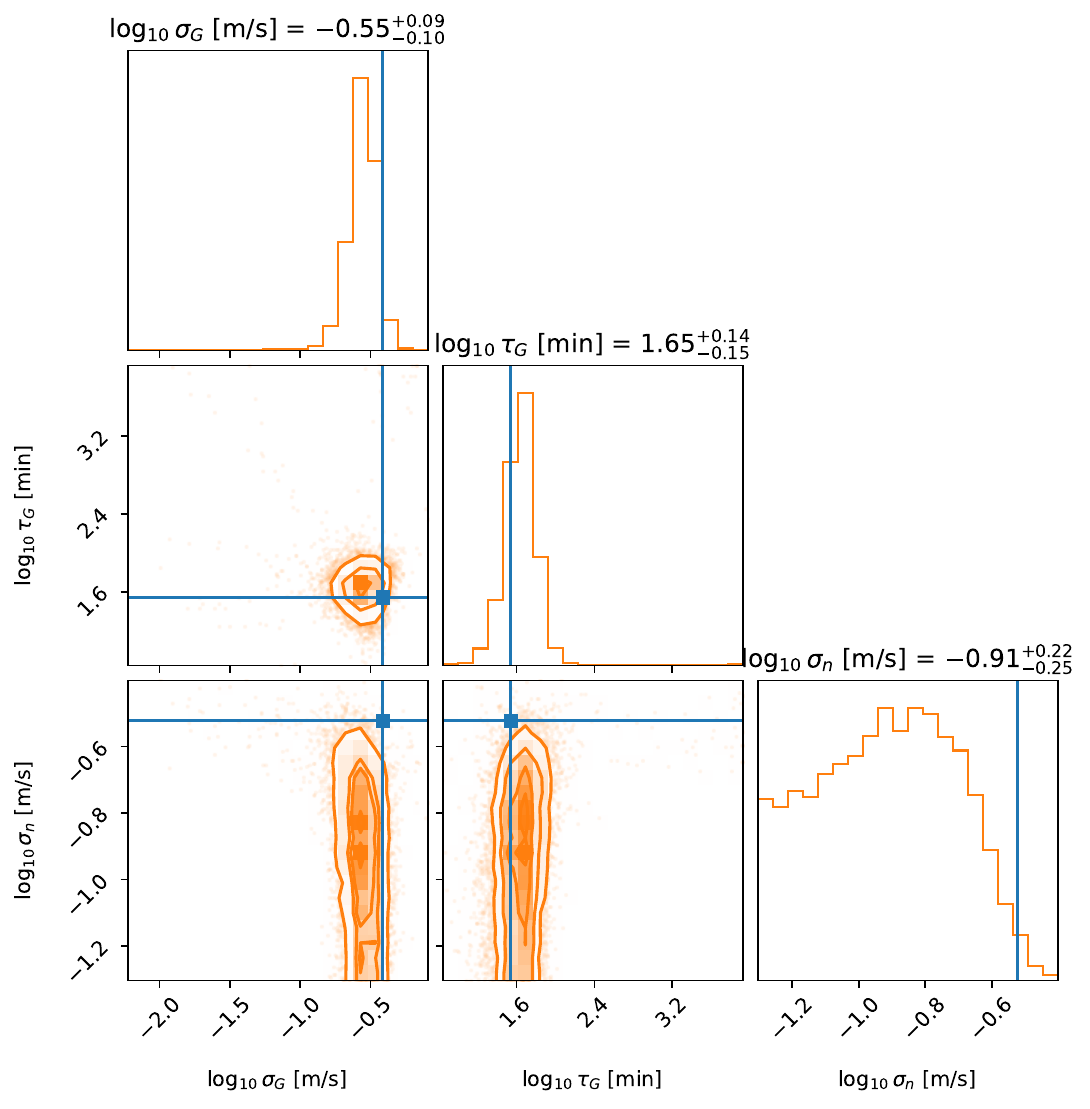}
    \caption{Corner plot for the one-aperiodic model fit to the first night of de-trended HARPS-N RVs. The blue lines represent the predicted values from the scaling relations.}
    \label{fig:night 1 corner plot}
\end{figure}

\begin{figure}
    \centering
    \includegraphics[width=0.48\textwidth]{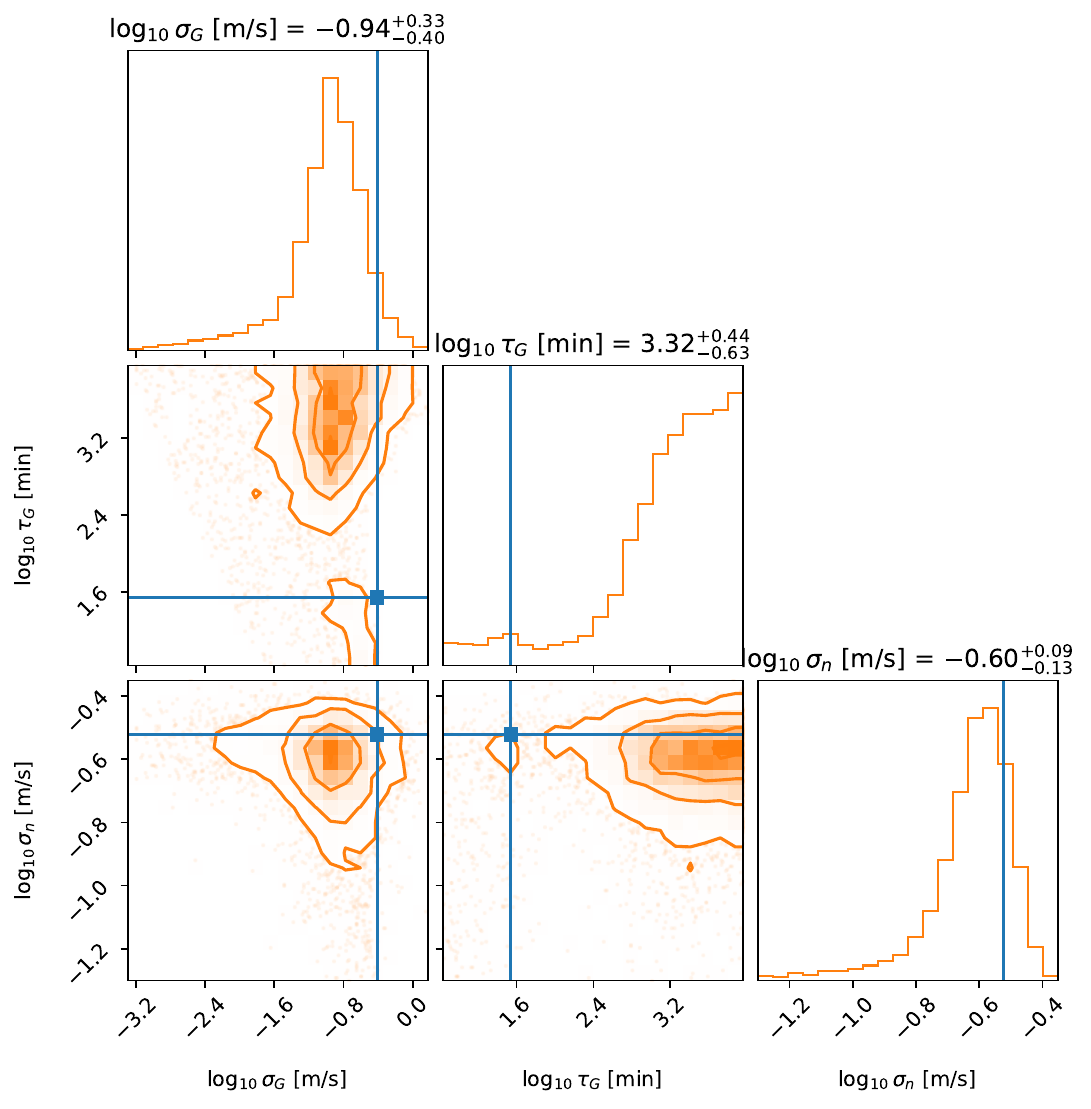}
    \caption{Corner plot for the one-aperiodic model fit to the second night of de-trended HARPS-N RVs. The blue lines represent the predicted values from the scaling relations.}
    \label{fig:night 2 corner plots}
\end{figure}

\begin{figure}
    \centering
    \includegraphics[width=0.48\textwidth]{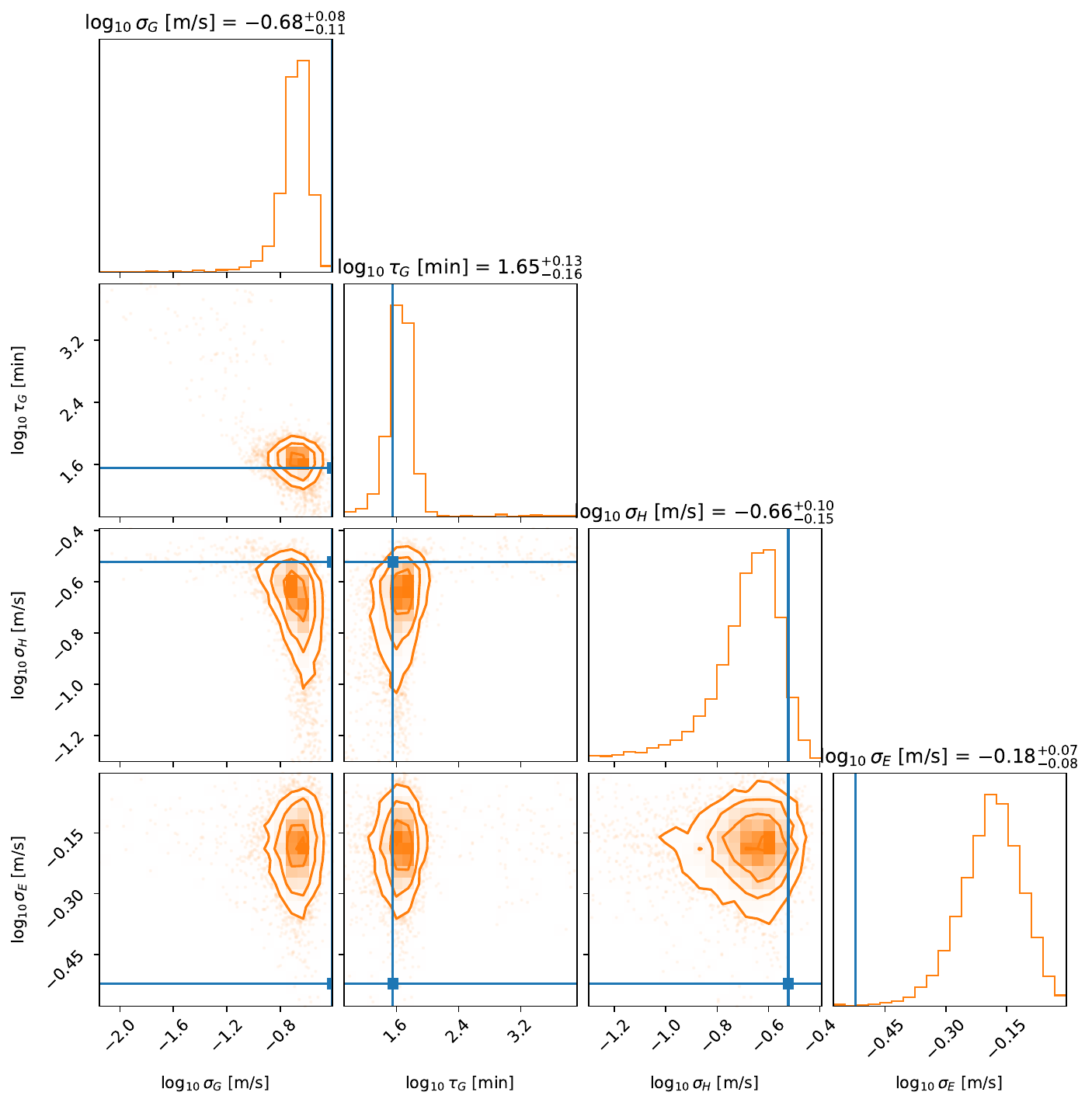}
    \caption{Corner plot for the one-aperiodic model fit to the combined HARPS-N and EXPRES RVs.The blue lines represent the predicted values from the scaling relations.}
    \label{fig:H+E}
\end{figure}

\section {High-low-high RMS measurements}
\label{sec:app_RMS}
\begin{figure}
    \centering
    \includegraphics[width=\linewidth]{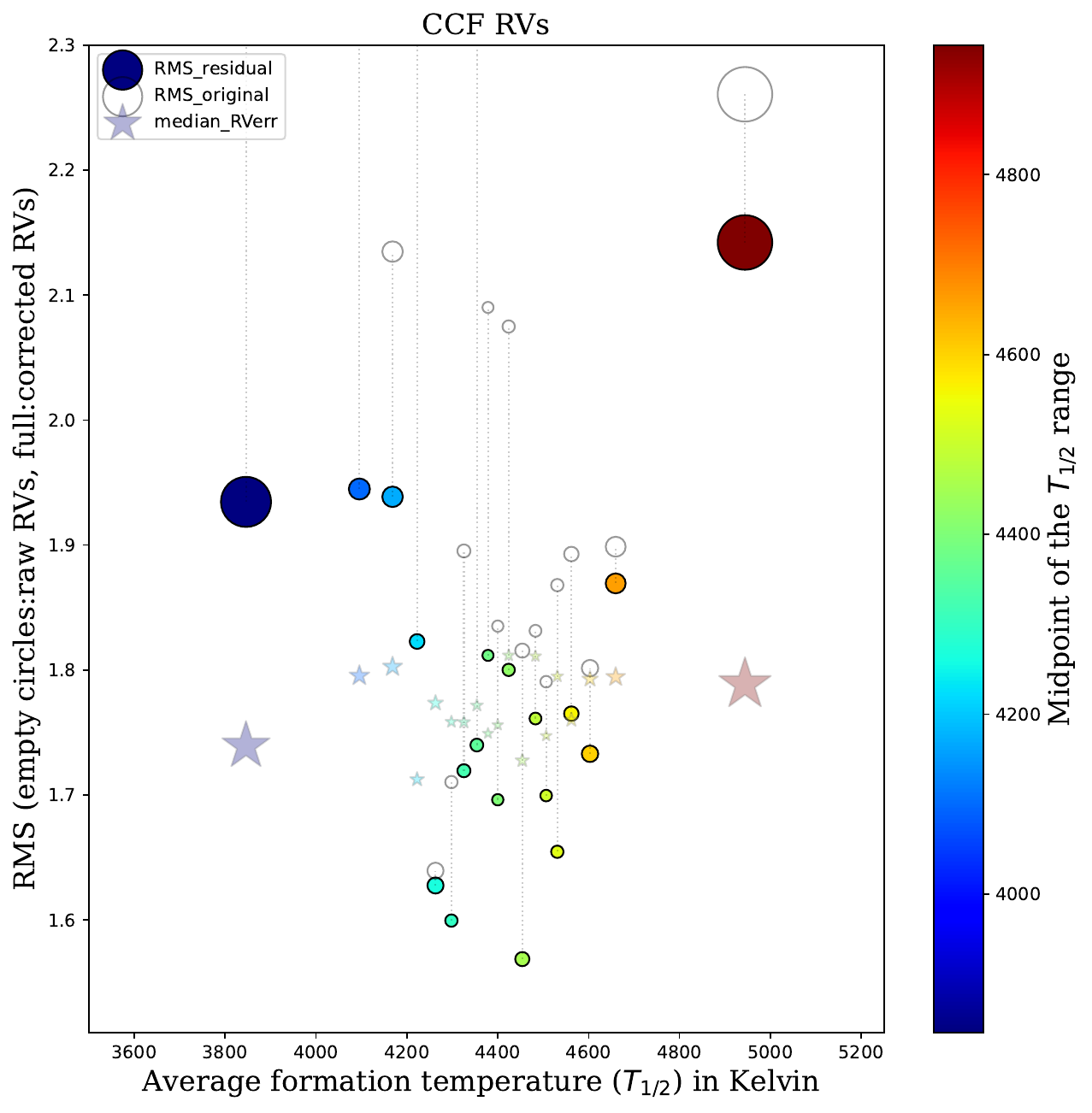}
    \caption{ The RMS values for the residual \textcolor{black}{$ {\rm CCF}_{T}$} RVs extracted from 20 distinct temperature ranges are displayed. These temperature bins are chosen to ensure that the RV content and variance are balanced. Faded stars represent the median RV error for each bin. An obvious high-low-high pattern emerges as we transition from the coolest to the hottest temperature bin. The semi-transparent circles represent the RMS values for the \textcolor{black}{$ {\rm CCF}_{T}$} RVs extracted from individual temperature ranges, before being corrected for the instrumental systematics. The marker sizes represent the sizes of the temperature bin. }
    \label{fig:app_RMS}
\end{figure}



\bsp	
\label{lastpage}
\end{document}